\documentclass[12pt,english]{article}
\usepackage[english]{babel}
\usepackage{amsmath,amssymb,amscd}
\usepackage{amsfonts}
\usepackage{indentfirst}
\usepackage{graphicx}

\makeatletter
\renewcommand\section{\@startsection {section}{1}{\z@}%
                                   {-5.5ex \@plus -1ex \@minus -.2ex}
                                   {2.3ex \@plus.2ex}%
                                   {\normalfont\large\bfseries}}
\renewcommand\subsection{\@startsection{subsection}{2}{\z@}%
                                     {-3.25ex\@plus -1ex \@minus -.2ex}%
                                     {1.5ex \@plus .2ex}%
                                     {\normalfont\bfseries}}

\newcommand\subsub[1]{
\bigskip

\noindent{\underline{\it #1}}
\smallskip}

\numberwithin{equation}{section}

\makeatother

\newcommand{\ba}{\begin{align}}

\newcommand{\bea}{\begin{eqnarray}}
\newcommand{\eea}{\end{eqnarray}}
\newcommand{\be}{\begin{equation}}
\newcommand{\ee}{\end{equation}}

\newcommand{\del}{\partial}

\newcommand{\hf}{\frac{1}{2}}

\newcommand{\Z}{{\mathbb Z}}
\newcommand{\R}{{\mathbb R}}
\newcommand{\C}{{\mathbb C}}

\renewcommand{\P}{{\mathbb P}}

\def\Tr{{\rm Tr}}

\def\la{\lambda}

\def\w{\omega}

\newcommand{\cB}{{\cal B }}
\newcommand{\cW}{{\cal W }}            
\newcommand{\cM}{{\cal M }}            
\newcommand{\cF}{{\cal F }}            

\newcommand{\cL}{{\cal L }}                
\newcommand{\cO}{{\cal O }}            
\newcommand{\cH}{{\cal H }}            
\newcommand{\cA}{{\cal A }}            

\newcommand{\cN}{{\cal N }}            
    
\newcommand{\cD}{{\cal D }} 
\newcommand{\cV}{{\cal V }}    
   
\newcommand{\B}{{\cal B}} 
\renewcommand{\d}{{\partial}}

\def\wt{\widetilde}

\begin{document}

\sloppy

\baselineskip=18pt

\begin{flushright}
\begin{tabular}{l}
ITFA-2008-42\\
BONN-TH-2008-13\\
 [.3in]
\end{tabular}
\end{flushright}

\begin{center}
{\Large \sc Quantum Curves and $\cD$-Modules \\[13mm]}

Robbert Dijkgraaf,$\!{}^{1,2}$ Lotte Hollands,$\!{}^1$ and 
Piotr Su{\l}kowski${}^{\,3,4}$\\[7mm]

\emph{$^1$\!\! Institute for Theoretical Physics, and \,$^2$\!\! KdV Institute
  for Mathematics,} \\ 
\emph{University of Amsterdam, Valckenierstraat
  65, 1018 XE Amsterdam, The Netherlands} \\[4mm]
\emph{$^3$\!\!  Physikalisches Institut der Universit{\"a}t Bonn and Bethe Center for Theoretical Physics,} \\
\emph{Nussallee 12, 53115 Bonn, Germany} \\  [4mm]
\emph{$^4$\!\! So{\l}tan Institute for Nuclear Studies, ul. Ho\.za 69, 00-681 Warsaw, Poland} \\

\end{center}

\vspace{1cm}

\centerline{\bf Abstract}
\smallskip
\begin{quote}

In this article we continue our study of chiral fermions on a quantum
curve. This system is embedded in string theory as an I-brane
configuration, which consists of D4 and D6-branes intersecting along a
holomorphic curve in a complex surface, together with a
$B$-field. Mathematically, it is described by a holonomic
$\cD$-module. Here we focus on spectral curves, which play a prominent role in the
theory of (quantum) integrable hierarchies. We show how to associate a
quantum state to the I-brane system, and subsequently how to compute
quantum invariants. As a first example, this yields an insightful
formulation of (double scaled as well as general Hermitian) matrix
models. Secondly, we formulate $c=1$ string theory in this language.
Finally, our formalism elegantly reconstructs the complete
dual Nekrasov-Okounkov partition function from a quantum
Seiberg-Witten curve.

\end{quote}

\newpage

\tableofcontents

\newpage 

\section{Introduction}

The study of two-dimensional conformal field theories on Riemann
surfaces has had many fruitful applications in physics and
mathematics. In many ways the field theory of a chiral free fermion is
the most important and instructive example. In that case one considers
a Riemann surface or algebraic curve $\Sigma$ together with a line
bundle ${\cal L}$ that comes equipped with connection $A$. The free
fermion partition function computes the determinant of the twisted
Dirac operator $\overline\partial_A$ coupled to the line bundle $\cal
L$. This determinant has many interesting properties, {\it e.g.}\ the
dependence of this determinant on the connection $A$ is captured by
the Jacobi theta-function.
It is known for a long time that these chiral determinants are closely
related to integrable hierarchies of KP-type \cite{sato, DJKM,
  segalwilson}. In the simplest case 
this relation arises as follows. One picks a point $P\in \Sigma$ on
the curve together with a local trivialization $e^t$ of the line bundle
around $P$. The ratio with respect to a reference connection $A_0$
$$
\tau(t) = {\det \overline\partial_{A} \over \det \overline\partial_{A_0}}
$$
then becomes a so-called tau-function of the KP-hierarchy.
In the Hamiltonian formulation one associates a state $|\cW\rangle$
in the fermionic Fock space $\cal F$ to the line bundle on $\Sigma -
P$. In the semi-infinite wedge representation of the Fock space this
state can be considered as the wedge product of a basis that spans the
space of holomorphic sections
$$
\cW=H^0(\Sigma-P, {\cal L}).
$$
Similarly, a coherent state $|t\rangle$ is associated to the local
trivialisation around $P$. Combining these two ingredients the
tau-function can be written as
\be
\tau(t) = \langle t |\cW \rangle.
\label{taufunction}
\ee
With the advent of matrix models it became clear that string theory
can also give rise to solutions of KP-type of the form
(\ref{taufunction}). More recently this connection to integrable
hierarchies has been reformulated and generalized through the methods
of topological strings \cite{adkmv,dhsv}. These string theory
solutions are similar, but 
not equivalent, to the familiar geometric solutions coming from CFT
that are sketched above. In particular the relevant Fock space state
$|\cW\rangle$ does not have a purely geometric interpretation as
generated by a space of sections over a curve. Yet, in the string
theory setting an algebraic curve $\Sigma$ does appear. (Here it
should be stressed that this curve is not a string world-sheet, but
should be considered as (part of) the target space geometry.) But in
this case there is an extra parameter: the string coupling constant
$\lambda$. Only in the genus zero or classical limit $\lambda \to 0$ a
geometric curve arises.
There have been many indications that $\lambda$ should be interpreted
as some form of non-commutative deformation of the underlying
algebraic curve. In the simplest cases $\Sigma$ appears as an affine
rational curve given by a relation of the form
$$
F(x,y)=0,
$$
in the complex two-plane $\C^2$, with a (local) parametrization
$$
x=p(z), \qquad y= q(z),
$$
with $p,q$ polynomials. Of course, $p$ and $q$ commute: $[p,q]=0$.
However, the string-type solutions with $\lambda \not= 0$ are
characterized by quantities $P$ and $Q$ that no longer commute but instead
satisfy the canonical commutation relation
$$
[P,Q]=\lambda.
$$
In this case clearly $P$, $Q$ cannot be polynomials, but are
represented as differential operators, {\it i.e.} polynomials in $z$
and $\partial_z$.
As we will point out in this paper a suitable concept to frame these
solutions is a $\cD$-module. Instead of classical curve in
the $(x,y)$-plane,
we should think of a {\it quantum curve} as an analogue in the
non-commutative plane $[x,y]=\lambda$. If we interpret
$$
y = - \lambda {\partial \over \partial x},
$$
one can identify such a quantum curve as a holonomic $\cD$-module
$\cW$ for the algebra $\cD$ of differential operators in $x$.
Now there is a straightforward way in which such a $\cD$-module gives
rise to a solution of the KP-hierarchy. By definition $\cW$ carries an
action of both $x$ and $\partial_x$. However we are free to ignore the
second action, which leaves us with the structure of an $\cO$-module,
$\cO$ being the algebra of functions in $x$. By applying the
infinite-wedge construction to the module $\cW$ we obtain in the usual
way a state $|\cW\rangle$ in the fermion Fock space. Roughly speaking,
$\cW$ can be considered as the space of local sections that can be
continued as sections of a (non-commutative) $\cD$-module, instead
of sections of a line bundle over a curve.
This set-up can be generalized in many ways and in this fashion several
constructions in topological string theory, matrix models and
integrable hierarchies can be connected. It is the purpose of this
paper to explain the connections between these familiar ingredients
from the $\cD$-module perspective. \\

\noindent This paper is structured as follows: 

In section~\ref{sec:formalism} we introduce our notion of a quantum curve and provide a construction of a tau-function associated to it. 
This tau-function arises, in an  appropriate sense, from a quantization of
the Krichever correspondence described in \S~\ref{ssec:Krichever}.
The physical system relevant for this quantization consists of an intersecting
brane configuration with $B$-field in string theory. It is introduced in \cite{dhsv} and reviewed in \S~\ref{ssec-Ibrane}. 
The D4 and D6-branes wrap an affine complex curve $\Sigma$ that is embedded in a complex symplectic
plane. Endpoints of the strings stretched between D4 and D6-branes appear as fermionic modes on
$\Sigma$, which are quantized by the $B$-field. This turns the
chiral fermions into sections of a so-called $\cD$-module. 
In \S~\ref{ssec-Ibrane} we explain in which sense a $\cD$-module
quantizes the spectral curve and in \S~\ref{ssec-prescription} we
discuss how one can associate a fermionic state $| \cW \rangle$, and
thus a tau-function,  to such a $\cD$-module. 

In sections~\ref{sec:matrixmodels}, \ref{sec:c=1} and \ref{sec:SW} we
analyse three physical systems in which above quantization is
realized: respectively
matrix models, $c=1$ string theory and $\cN=2$ supersymmetric gauge
theories. We find that a quantization of the underlying classical curve yields a differential system that determines the corresponding partition functions. In other words, we see how our formalism in section~\ref{sec:formalism} gives a unifying picture of these topics in terms of a underlying quantum curve.

\section{Quantum curves and invariants}\label{sec:formalism} 

The main object of interest in this paper is a chiral fermion field
living on a holomorphic quantum curve. This set-up is embedded in
string theory as a configuration of D4 and D6-branes that intersect
along a classical curve $\Sigma$. Turning on a B-field on the D6-brane
quantizes the curve $\Sigma$. As was shown in \cite{dhsv} this
intersecting brane configuration is closely related to topological
string theory, supersymmetric gauge theory and matrix models.     
More precisely, it relates to the topological B-model on non-compact
Calabi-Yau backgrounds of the form 
\begin{align*}
X_{\Sigma}:~ uv - F(z,w) = 0,
\end{align*}
that is modeled on an affine curve
%
$\Sigma$ defined by the equation $ F(x,y) =0$.
%
The topological string partition function admits an
expansion 
\begin{align*}
Z_{\mathrm{top}}(t,\la) = \exp \left( \sum \la^{2g-2} \cF_g \right)
\end{align*}
in the topological coupling constant $\la$, whose classical
contribution $\cF_0$ captures the complex periods 
\begin{align*}
X_i = \int_{A_i} \Omega, \quad \partial_i \cF_0 = \int_{B_i} \Omega
\end{align*}
of $X_{\Sigma}$, while the semi-classical contribution $\cF_1$ is known
to compute a chiral determinant 
\begin{align}\label{eq:chiraldet}
\exp \cF_1 = \mathrm{det} ~\overline{\partial}_{\Sigma}
\end{align}
on $\Sigma$. All higher order $\cF_g$'s give quantum corrections
to these results. 

As we alluded to in the introduction, the chiral determinant
(\ref{eq:chiraldet}) has an elegant interpretation in terms of certain
geometric solutions of the KP hierarchy, which is known as the
Krichever correspondence. In this context the chiral determinant is known
as a tau-function. On the other hand, the total topological string
partition function is also known to represent a tau-function of a KP
hierarchy, though in this case it doesn't have a similar geometric
interpretation. The aim of this section is to propose a quantum analog
of the Krichever correspondence, starting from a quantum curve. We
conjecture that this prescription computes the all-genus topological
string partition function.        

In this section we start by reviewing the Krichever correspondence. We
continue by reviewing the intersecting brane system and explain what
we mean by a quantum curve. In the last subsection we line out our
prescription to obtain invariants from such a quantum curve.

\subsection{Krichever correspondence}   \label{ssec:Krichever}

In this section we review the geometric Krichever
correspondence that underlies the genus 1 free energy $\cF_1$ of the
topological string. 
In the simplest scenario we start with a Riemann surface $\Sigma$ with
a single puncture $P$. We study a chiral fermion field 
\begin{align*}
\psi(z) = \sum_{r \in \Z + 1/2} \psi_r z^{-r-1/2},\qquad \quad  \{ \psi^{\dag}_r, \psi_s \} = \delta_{r+s,0}
\end{align*}
on $\Sigma$ which is coupled to a line bundle $\cL$. 
The Hilbert space $\cH$ of this fermion field 
is built by acting on the Dirac vacuum $|0\rangle$ with the fermionic
modes $\psi_r$ and $\psi^{\dag}_r$.
The boundary conditions near $P$, \emph{i.e.} the choice of local
coordinates and a choice of local frame of $\cL$ near
$P$, are encoded in a coherent state 
$$
\langle t | = \langle 0 | e^{\sum t_n  \alpha_n},
$$
where $\alpha_n = \sum_r : \psi_r^{\dag} \psi_{n-r}:$ are the bosonized
modes. The partition function of the fermion field sweeps out a state $| \cW
\rangle$ in the Hilbert space $\cH$, and for a given choice $t$ of boundary conditions
it reads  
\begin{align}\label{eqn:taufunction}
\tau(t) = \langle t | \cW \rangle. 
\end{align}

The Krichever correspondence tells us precisely how to find the state
$| \cW \rangle$. Choosing $z^{-1}$ as a local coordinate around $P$, we
define a subspace 
\begin{align}
\cW \equiv H^0(\Sigma - P, \cL) \subset \C[z] \oplus \C[[z^{-1}]].
\end{align}
By picking a semi-infinite basis $w_n = z^n(1 + \cO(z^{-1}))$ of this
subspace, it can be quantized into a fermionic state
\begin{align}
| \cW \rangle = w_1 \wedge w_2 \wedge w_3 \wedge \ldots \in \cH.
\end{align}
on which the fermionic modes act as 
\begin{align*}
\psi_r = \frac{\partial}{\partial z^{-r+1/2}}, \quad \psi^{\dag}_r =
z^{r-1/2} \wedge .
\end{align*}
Any state $|\cW \rangle$ that we obtain in this way looks like 
\begin{align}
| \cW \rangle = g |0 \rangle, \quad \textrm{where}~
 g = \exp \left( \sum c_{nm} \psi_n \psi_m^{\dag} \right) \in Gl(\infty).
\end{align} 
All states of the above form parametrize an infinite
Grassmannian, which is well-known to give an elegant geometric
formulation of the KP integrable hierarchy. (A more
detailed review of these issues can be found in Appendix A, which will
be useful later on.)  
  
Important for now is that although one can associate a tau function to
any element $| \cW \rangle $ in the Grassmannian, as in
equation~(\ref{eqn:taufunction}), only a dense subset of subspaces
$\cW$ in the infinite Grassmannian of measure 0 allows for a
geometric Krichever interpretation.    
This subset can be characterized as follows. Basically, a subspace
$\cW$ has a geometric origin when there is an algebra $\cA$ such that
\begin{align}\label{eq:clasalgebra}
\cA \cdot \cW \subset \cW,
\end{align}
with $\cA$ being non-trivial, \emph{i.e.} $\cA \neq \C$. In this situation 
the underlying curve $\Sigma$ can be defined in terms of its spectrum
$\cA = H^0(\Sigma- P, \cL)$ and the tau-function
(\ref{eqn:taufunction}) has an interpretation as a fermionic
determinant $\textrm{det}~ \overline{\partial}_{\Sigma}$. 

\subsection{B-field, $\cD$-modules and quantum curve}     \label{ssec-Ibrane}

Our motivation for quantizing the Krichever correspondence
comes from \cite{dhsv}, where it is argued that
topological string theory is related by a chain of dualities to a system of intersecting D4 and D6
branes in type IIA with a background $B$-field. We now review some aspects of this relation.

\subsub{I-brane configuration}

In the intersecting brane set-up a central role is played by a
holomorphically embedded curve $\Sigma \subset \B$, given by the
equation      
$$
\Sigma:\ F(z,w)=0,
$$
where $\B = \C \times \C$ (or possibly with either $\C$ replaced by
$\C^*$) is parametrized by complex coordinates $(z,w)$.
We consider this curve in the type IIA background 
\be
\hbox{(IIA)}\quad \R^3 \times \B \times \R^2\times S^1,   \label{I-setup}
\ee
and place a D4-brane wrapping $\R^3 \times \Sigma$ and a D6-brane
wrapping  $\B \times \R^2\times S^1$. These branes intersect over
$\Sigma$. Fermions on $\Sigma$ are realized by massless modes of the
4-6 strings. The supersymmetry of the system ensures holomorphicity. The supersymmetries act trivially on the chiral fermions, which constitute a topological subsector of the complete system. 

Non-commutativity in this set-up is introduced by turning on a
constant $B$-field along $\B$, with holomorphic part 
\be
B = \frac{1}{\lambda} dz\wedge dw   \label{B-lambda}.
\ee
It is realized on the worldvolume of the D6-brane.

\begin{figure}[h!]
\begin{center}   \label{fig1}
\includegraphics[width=7cm]{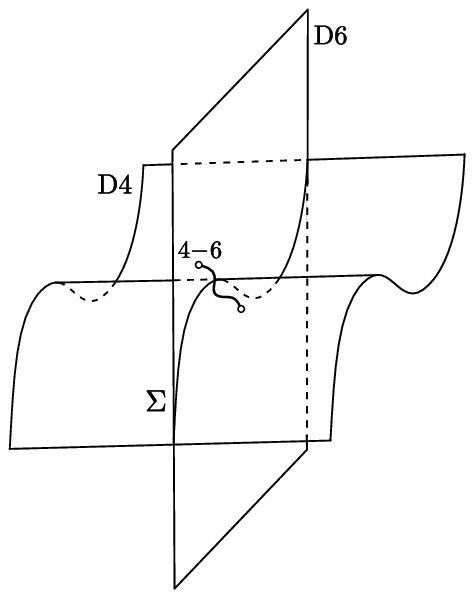}\\[5mm]
\parbox{12cm}{\small \bf Fig.\ 1: \it The I-brane configuration. A
  D4-brane intersects with a D6-brane along a curve $\Sigma$. The 4-6
  string degrees of freedom show up as free fermions on $\Sigma$.}
\end{center} 
\end{figure}

By a chain of dualities presented in~\cite{dhsv} this I-brane configuration relates
to the background  
\be
\hbox{(IIA)}\quad \R^3 \times \wt X \times S^1,  \label{DT-setup}
\ee
with a $D6$-brane wrapping $\wt X \times S^1$. This setup is appropriate
for a computation of Donaldson-Thomas invariants $DT(n,d)$, physically
interpreted as BPS bound states of  $n \in H_0(\wt X,\Z)\cong \Z$
D0-branes and  $d \in H_2(\wt X,\Z)$ 
D2-branes to the D6-brane. The generating function of these invariants 
\be
\label{DT}
Z_{\mathrm{qu}}(t,\la) =  \sum_{n,d} DT(n,d) \, e^{-n\la} e^{d \cdot t}
\ee
is closely related to the A-model topological string partition
function on the toric manifold $\wt X$ with the complexified K\"ahler
class $t \in H^2(\wt X)$ 
\be
Z_{\mathrm{top}}(t,\la)  =
\exp \left(-{t^3\over 6 \la^2} - {1\over 24} t \cdot c_2(\wt
 X)\right) Z_{\mathrm{qu}}(t,\la).   \label{Z-top}
\ee

Following the duality chain mentioned above, the (holomorphic) parameter
$\lambda$ which initially specified a value of the $B$-field
(\ref{B-lambda}) acquires an interpretation of the topological string
coupling constant $\lambda$ \cite{dhsv}. After summing over 
all bound states with $p$ D4-branes as well, while weighting their
contribution with a potential $\xi$, the partition function of the
final configuration reads  
\begin{align}\label{fermionpartfunc}
Z_{\mathrm{I}}(\xi,t,\la) &= 
\sum_{p \in H^2(\wt X,\Z)} e^{p\xi}Z_{\mathrm{top}}(t+p\la,\la).
\end{align}
We identify this partition function with the I-brane partition
function of the initial configuration (\ref{I-setup}).  

The above system can also be easily related to the supersymmetric
gauge theories leading to a system of D4-branes spanned between
NS5-branes. As shown by Witten \cite{M-4d}, such a configuration engineers
$\cN=2$ supersymmetric  gauge theories. We will elucidate this relation in much detail
in section \ref{sec:SW}.

\subsub{$B$-field}\label{sec:Dmods}

The B-field quantizes the fermions on $\Sigma$.
Let us first repeat the general arguments of \cite{dhsv}. 
The algebra $\cA$ of open 6-6 strings on the D6-brane describes 
the interaction (as illustrated in figure~2)
\be
\cA \otimes \cA \to \cA,
\ee
and is explicitly non-commutative in the presence of a $B$-field. The
$B$-field 
introduces a gauge field $A$ on the D6-brane that couples to the open
strings and quantizes the algebra of zero-modes of those strings
\cite{sw-noncom,langlands}. With a $B$-field given by
\begin{align}
B = \frac{1}{\la} dz \wedge dw
\end{align}
the non-commutativity parameter is $\la$. The complex coordinates
$z$ and $w$ become non-commutative operators obeying
\begin{align}
[z,w] = \la.   \label{I-Weyl}
\end{align}
In case $\cB = \C \times \C$ we can identify this algebra
with the Weyl algebra of differential operators
\bea
\cA \cong \cD_\C = \langle z, \la \d_z \rangle.
\eea
%

When we add to this system a D4-brane intersecting the D6-brane
along the curve $\Sigma$, a 6-6 string acting on a 4-6 string can produce another 4-6 string (see figure~2)
\bea
\cA \times \cM \to \cM.
\eea
This action endows the space of 4-6 open strings $\cM$ with the structure of a module over the algebra $\cA$ of 6-6 strings. 
Modules for the algebra of differential operators are called
$\cD$-modules.

To conclude, in the presence of a background flux, the chiral fermions on the I-brane should no longer be regarded as sections of the spin bundle $K^{1/2}$. Instead they should be viewed as sections of a $\cD$-module. 

In the context of string theory it is worth stressing the range of parameters $\alpha'$ and $\la$ in which $\cD$-module description is valid. The string coupling $\la$, which enters in the B-field flux as $B = \frac{1}{\la} dz \wedge dw$, plays an important role as quantization parameter. From the $\cD$-module point of view there seems to be no restriction on $\la$, so one might hope that the $\cD$-module even captures non-perturbative information.  However, in a particular system under consideration some restrictions on the values of $\la$ could arise that are related to the radius of convergence of the partition function. Although we do make some additional remarks in Section~\ref{sec:matrixmodels} and in Section~\ref{sec:discussion}, we do not study these issues in this paper.

On the other hand, the string scale $\alpha'$ does not play a fundamental role in the $\cD$-module. The $\cD$-module describes the topological sector of the intersecting brane configuration, which is realized in terms of massless modes of the I-brane system. Therefore the $\cD$-module description is valid only in the regime where $\alpha'$ is small (so that no massive modes interfere with our description). The most interesting case is of course when it is non-zero, as it provides a normalization factor for the worldsheet instanton contributions to the open 4-6 strings in the I-brane partition function (\ref{fermionpartfunc}). Section~\ref{sec:SW} clarifies this with an example.

\begin{figure}[h!]
\begin{center}   \label{fig-module}
\includegraphics[width=10.5cm]{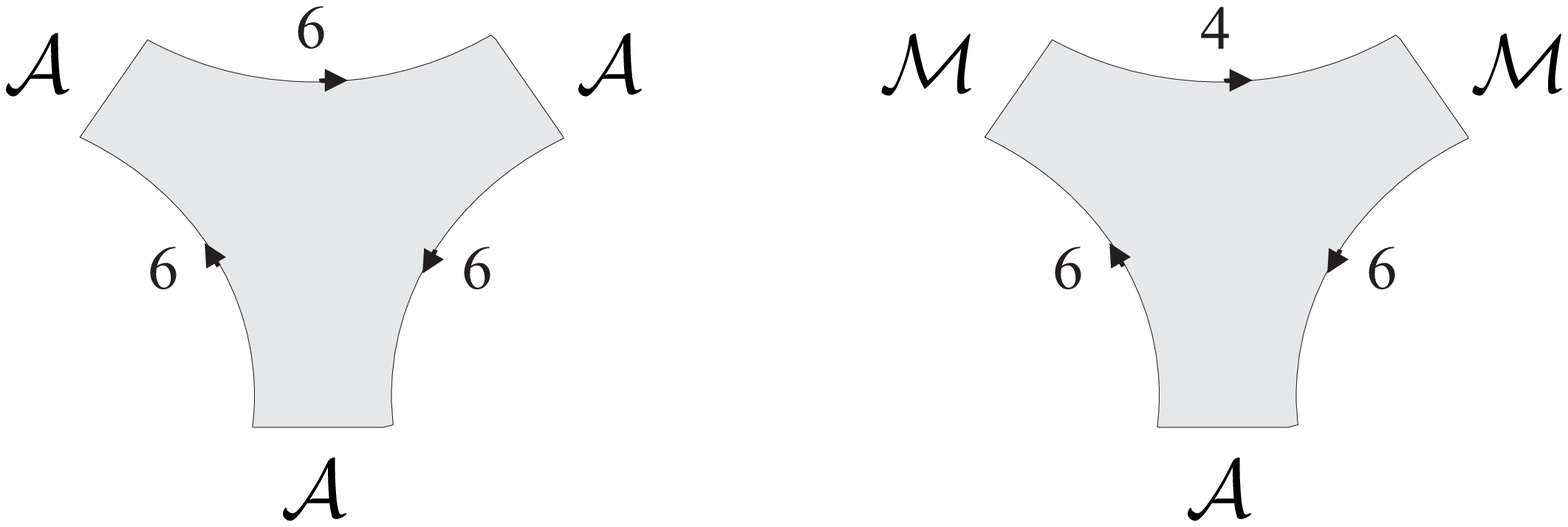}\\[5mm]
\parbox{12cm}{\small \bf Fig.\ 2: \it  The algebra $\cA$ of functions on
  $\Sigma$ acts on the module $\cM$ of free fermions. In the presence
  of a $B$-field the algebra $\cA$ may be represented as a
  differential algebra, so that $\cM$ becomes a $\cD$-module.}
\end{center} 
\end{figure}

\subsub{$\cD$-modules and quantum curves}

We here introduce basic facts concerning $\cD$-modules and explain why they naturally describe I-branes.
More details concerning theory of $\cD$-modules can be found in Appendix~\ref{sec:dmods}. 

$\cD$-modules are defined as modules for the algebra of differential operators $\cD$. 
In this paper we are interested in $\cD$-modules for the Weyl
algebra $\cD = \langle z, \partial_z \rangle$. These are affine $\cD$-modules
of rank 1 and can represented as
$$ \cM = \frac{\cD}{\cD \cdot P },$$
where $P$ is a linear differential operator $P = \sum_i a_i(z) \partial^i_z.$
The module $\cM$ therefore captures solutions to the differential equation 
\be
P\Psi = 0,  
\ee 
where $\Psi$ takes values in some function space $\cV$, for example the
algebra $\cO_{\C}$ of holomorphic functions on the complex plane $\C$. 
%
%
%
%
$\cD$-modules of rank 1 are cyclic, {\it i.e.} they are generated by a single element $\Psi
\in\cM$, and so are of the form 
\be
\cM = \{D\Psi : D\in\cD\}.   \label{eqn:DPsi}
\ee

To be more precise, $\cD$-modules generated by the $B$-field
(\ref{B-lambda}) depend on $\la$ and are known as  
$\cD_{\la}$-modules \cite{arinkin} (when $\la$ is considered as a
formal variable). Since all the differential modules we consider are
$\cD_{\la}$-modules, we often omit the subscript $\la$.

The $\cD$-module structure $\cD \cdot \cM \subset \cM$
gives a quantization of the semi-classical description in
equation~(\ref{eq:clasalgebra}). In particular, the rank 1 $\cD$-module 
\begin{align}
\cM =  \frac{\cD}{\cD  \cdot P(z)} 
\end{align}
is a quantization of the module
\begin{align}
\cW =  \frac{\cO_{\{z,w\}}}{\cO_{\{z,w\}}  \cdot F(z,w)} 
\end{align}
of functions on the curve defined by $F(z,w)=0$. We therefore refer to
the underlying differential equation $P(z)=0$ as a quantum curve. The
I-brane set-up will obviously provide us with a rank 1 $\cD$-module that represents a quantization of the I-brane curve $\Sigma$.\footnote{As a side remark notice that holonomic D-modules of dimension higher than 2 cannot be embedded in the 10 dimensions of string theory. Holonomic D-modules of dimension 2 are not related to the type II Calabi-Yau compactifications that we study in this paper, but could play a role in 4-dimensional Calabi-Yau compactifications in F-theory.}

Our notion of a quantum curve agrees with a notion of quantum spectral
curves in the theory of (Hitchin) integrable systems. We discuss this
relation shortly in Appendix~\ref{sec:quantuminthier}.
Here we just give some examples of $\cD$-modules and their
interpretation in terms of quantum curves. 

\subsub{Examples}
\vspace{2mm}

\noindent 1)~
Take a linear partial differential operator on $\C$, for example 
\begin{align}
P = \la z \partial_z - 1.   \label{Pexmp}
\end{align}
The differential equation $P \Psi=0$ is solved by
$\Psi(z) = z^{1/\la}$. So according to (\ref{eqn:DPsi}) the corresponding
$\cD$-module can be represented as
\begin{align}
\cM = \langle z,\la\partial_z \rangle \, z^{1/\la}.
\end{align}

There are many equivalent ways of writing this module. For example,
introducing $\wt \Psi = z\Psi$, 
the above differential equation is transformed into ${\wt P} {\wt
  \Psi} = 0$ with  
\be
{\wt P} = \la z \partial_z - \lambda - 1.
\ee
This new operator, as well as the solution to the new equation ${\wt
  \Psi} = z^{1+1/\la}$ look 
different than before. Nonetheless, they represent the same $\cD$-module
\be
\cM = \langle z,\la \partial_z \rangle \, z^{1+1/\la} = \langle
z,\la \partial_z \rangle \, z^{1/\la}. 
\ee
This simple example illustrates how the formalism of $\cD$-modules
allows to study solutions to partial differential equations  
independently of the way in which they are written.

An equivalent way to study $\cD$-modules is in terms of flat
connections  (see equation (\ref{eq:connectioncorrD-mod})). The flat
connection corresponding to $P$ is given by  
 \begin{align}
 \nabla_A = \la \partial_z dz - \frac{1}{z} dz, 
 \end{align}
and determines $\Psi(z)$ as a local flat section. It is a $\la$-deformation of the degree 1 spectral cover
\begin{align}
\Sigma: \quad  w = \frac{1}{z},  
\end{align}
with $z,w \in \C^*$, together with the (meromorphic) 1-form  
\begin{align}
A = \frac{1}{z} dz.
\end{align}
%

This example enters string theory as the deformed
conifold geometry describing the $c=1$ string. We will come back to it
in section~\ref{sec:c=1}. \\ 

\begin{figure}[h!]
\begin{center}   \label{fig2}
\includegraphics[width=4.5cm]{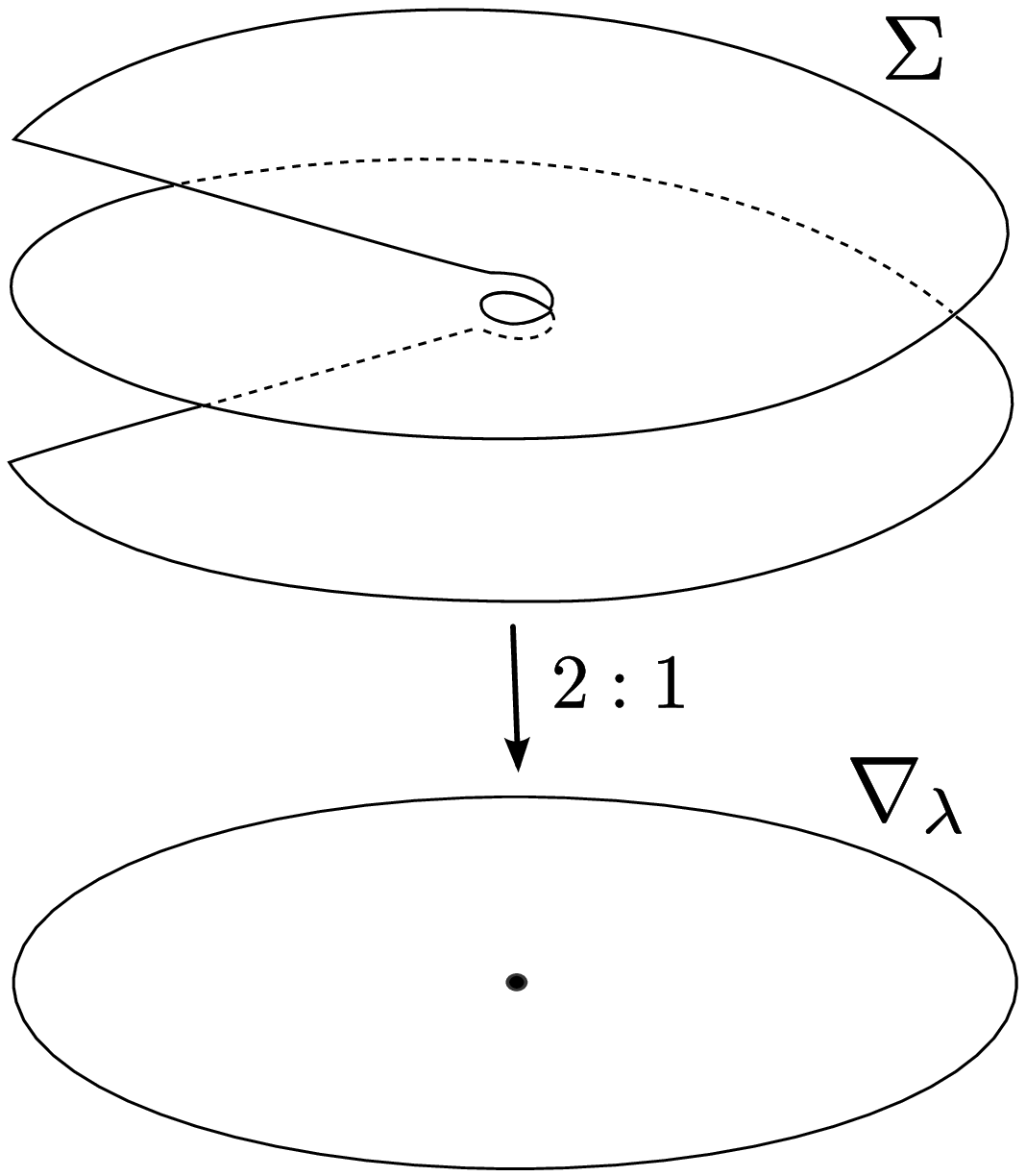}\\[5mm]
\parbox{12cm}{\small \bf Fig.\ 3: \it A second order differential
  operator $P$ in $\la \partial_z$ defines a rank 2 $\la$-connection
  $\nabla_{\la}$. The determinant of $\nabla_0$ determines a degree 2
  cover over $\C$ which is called the spectral curve $\Sigma$.}
\end{center} 
\end{figure}

\noindent 2)~
All the modules that we will study in this paper are over $\C$ or
$\C^*$. It is important that they may be of any rank though. Let us
therefore also give a rank two example on the complex plane $\C$. The 
second order differential equation
\begin{align}
P \Psi= (\la^2 \partial_z^2 - z)\Psi
\end{align}
can be written equivalently as a rank two differential system  
\begin{align}
P_{ij} \psi_j = 0, \quad \mbox{with}~ P_{ij}=  \left(\begin{array}{cc} 
   \la \partial_z & 0 \\ 0  & 
   \la \partial_z \end{array} \right) -  \left(\begin{array}{cc} 
  0 & 1 \\ z   & 
  0 \end{array} \right). 
\end{align}
Holomorphic solutions of this linear system are captured by the map 
\begin{align}
\cM = \frac{\cD^{\oplus 2}}{\cD^{\oplus 2} \, P_{ij}}
\rightarrow \cO_{\C}^{\oplus 2}
\end{align}
that sends the two generators $[(1,0)^t]$ and $[(0,1)^t]$ to two
independent (2-vector) solutions of $P \Psi = 0$. The corresponding
flat connection 
\begin{align}
\nabla_A = 
\partial_z dz - \frac{1}{\la}
\left( \begin{array}{cc} 0 & 1 \\ z &   0 \end{array} \right)  dz 
\end{align}
is a $\la$-deformation of the degree 2 spectral cover (illustrated in figure~3) 
\begin{align}
\Sigma: ~w^2 =  z,
\end{align}
with meromorphic 1-form $\eta =  w dz|_{\Sigma}$. Note that this
one-form pushes forward to the connection 1-form, or Higgs
field,
\begin{align}
A =  \left(\begin{array}{cc} 
  0 & 1 \\  z   & 
  0 \end{array} \right) dz 
\end{align}
in the basis $\{ dz, wdz \}$ of ramification 1-forms on the
$z$-plane. Indeed, local sections of $L$ push forward to local
sections generated by $1$ and $w$ on the $z$-plane. Now, $wdz \cdot 1
= w dz$ and $ w dz \cdot w = w^2 dz = z dz$ on $\Sigma$.

We will discuss the string theory interpretation of this $\cD$-module
in detail in section~\ref{sec:matrixmodels}. 

\subsection{Prescription}             \label{ssec-prescription}

Locally $\cD$-module on a general curve is just a system of linear
differential equations, that changes from patch to patch. It
is therefore natural to try to associate quantum invariants locally, and glue
them with the help of the $\cD$-module transformations (which are a
quantization of the usual coordinate transformations). This is exactly
the strategy we follow in the examples treated in the following sections.    

The simplest examples are deformations of affine curves of genus 0 with a
single marked point at infinity, \emph{i.e.} consisting of a single
patch near infinity. In these examples we associate a quantum invariant to the curve in the following way:

We start with a differential equation $P(z)=0$ representing the quantum
curve, where $z^{-1}$ is the coordinate near infinity. 
Solutions to the differential equation $P(z) \Psi(z) = 0$ form a module
$\cM$, which is in particular $\cO$-module. We call this $\cO$-module
$\cW$, and expect it to yield a subspace
$$\cW \subset \C((z^{-1})).$$ 
In that case we can, analogously as in the semi-classical case,
turn $\cW$ into a fermionic state and compute a tau-function.

Since the I-brane configuration provides a $\cD_{\la}$-module (in
contrast to a $\cD$-module) the resulting  I-brane fermionic state
$|\cW \rangle$ is a $\la$-deformation as well. We conjecture that its
determinant computes the all-genus topological string partition
function, when the appropiate quantum curve is chosen.    

In next sections we also discuss examples where we need
to glue two local patches. Since each local patch yields a subspace of
functions in the local variable, it is clear how the
glueing should work: we need to insert a Fourier-like operator
that relates the $\cD$-module on the first patch to the one on the
second patch.  As a result, the partition function
turns into a correlation function which contracts the two corresponding
fermionic states, with the insertion of the corresponding
Fourier-like operator. (This is similar as in \cite{adkmv}.)   

The above recipe doesn't tell us how to quantize a classical
curve in a specific physical set-up. This is a very hard question
in general. Moreover, it is not clear that the above prescription is
independent of the chosen covering by local patches.   
Rather then providing a general theory, in the following sections we analyze
several important examples of spectral curves in string theory, and determine
$\cD_{\la}$-modules that underlie their partition functions. 
Before we start with these examples though, let us explain the local
procedure in more detail in two simple cases.  


\subsub{Examples}

\noindent 1) \quad 
Let's first explain the rank 1
case, with a $\cD_{\la}$-module on $\C$ specified by the (meromorphic)
connection 
%
$\nabla_{A} = \partial_z - \frac{1}{\la} A(z)$
%
that may be trivialized as 
\begin{align*}
\nabla_A =  \partial_z - g_{\la}(z)^{-1}
(\partial_z g_{\la}(z)). 
\end{align*}
When $g_{\la}(z)$ is a holomorphic function on $\C$ that equals $g_{\la}(0)=1$ at $z=0$ -- in
the notation of equation~(\ref{Gplus}) this is
an element of $ \Gamma_+$ -- this represents a
pure gauge transformation on the disk, so that $\nabla_A$ corresponds to a
regular flat connection on $\C$. 

For any $g_{\la}(z)$ a fermionic section 
$\psi(z)$ of $\cL \otimes K^{1/2}$ may be written as  
\begin{align*}
\psi(z) = g_{\la}(z) \xi(z),
\end{align*}
where $\xi(z)$ is a section of $\cL \otimes K^{1/2}$ with trivial
connection $\partial_z$. Flat sections $\Psi(z)$ are defined by the
differential equation  
\begin{align*}
\left( \partial_z - \frac{1}{\la} A(z) \right) \Psi(z) = 0.
\end{align*}
They define a local trivialization of the bundle $\cL$ with
connection $\nabla_A$, and we will use them to translate the geometric
configuration into a quantum state. 

A flat section for the
trivial connection $\partial_z$ is given by $\Xi(z) = 1$. We 
associate the pair $(\C,\partial_z)$ to the ground state 
%
$|0 \rangle = z^0 \wedge z^1 \wedge z^2 \wedge \ldots$.
%
The gauge transformation $g_{\la}(z)$ maps the trivial solution 
$\Xi(z) = 1$ to $\Psi(z) = g_{\la}(z)$, which transforms
the vacuum into the fermionic state  
\begin{align*}
|\cW \rangle = g_{\la} |0 \rangle \equiv g_{\la}(z) z^0 \wedge g_{\la}(z) z^1 \wedge.
\end{align*}
%
%
%
In other words, we build the quantum state by acting with the
$\cD$-module generator $\Psi(z) = g_{\la}(z)$ on the vacuum
\begin{align*}
\cW = \cD_{\la} \cdot \Psi(z).
\end{align*}
The state $|\cW \rangle$ is just the second quantization of the $\cD_{\la}$-module $\cW$.  
This state is non-trivial only when $g_{\la}(z)$ is not a pure 
gauge transformation (which would correspond to a Krichever solution). In this situation the flat section diverges near $z=0$, corresponding
to a distorted geometry in this region. \\

\noindent 2) \quad A degree $n$ spectral curve $\Sigma$ is quantized as a
$\la$-connection of rank
$n$. This is equivalent to a $\cD_{\la}$-module $\cM$
that is generated by a single degree $n$ differential operator
$P$.
%
%
As an $\cO_{\C}$-module, though, $\cM$ is generated by an $n$-tuple
\begin{align*}
(\Psi(z), \del_z \Psi(z), \ldots, \del_z^{n-1} \Psi(z) ),  
\end{align*}
where $\Psi(z)$ is a solution of the differential equation $P \Psi =
0$. In other words, this blends an $n$-vector of solutions
to the linear differential system that the $\la$-connection defines. 
We will name this $\cO_{\C}$-module 
\begin{align*}
\cW = \cO_{\C} \cdot (\Psi(z), \del_z \Psi(z), \ldots, \del_z^{n-1}
\Psi(z) ) ~\subset~ \C((z^{-1}))
\end{align*}
(of course it contains the same elements as $\cM$).
This is the subspace we want to second quantize into a fermionic state $|\cW \rangle$. 

$P$ has $n$ independent solutions $\Psi_i$ that differ in their behaviour
at infinity. These solutions have an asymptotic expansion around $z =
\infty$ that contains a WKB-piece plus an asymptotic
expansion in $\la$, and should thus be interpreted as
perturbative solutions that live on the spectral cover. We suggest
 that the asymptotic expansion of any solution can be turned into a
fermionic state that captures the all-genus I-brane partition
function. This partition function thus depends on the choice of
boundary conditions near $z_{\infty}$. 

Some of the WKB-factors will be exponentially suppressed near
$z_{\infty}$, while
others grow exponentially. This depends on the  
specific region in this neighbourhood. The lines that characterize the
changing behaviour of the solutions $\Psi_i$ are called Stokes and
anti-Stokes rays. 
Boundary conditions at infinity specify the solution up to a Stokes
matrix: a solution that decays in that region can be added at no cost. 

This implies that the perturbative fermionic state we assign to a $\cD$-module
depends on the choice of boundary conditions. On the other hand, the
$\cD$-module itself is independent of 
any of these choices and thus in some sense contains non-perturbative
information and goes beyond the all-genus I-brane partition
function. This agrees with the discussion in
\cite{Maldacena:2004sn}. Nonetheless, the focus in this paper is on the perturbative information a $\cD$-module provides.




\section{Matrix model geometries}\label{sec:matrixmodels}

Hermitian one-matrix models with potential $W(x)= \sum_{j=0}^{d+1} u_j x^j$ are defined through the matrix integral  
\begin{align}\label{eq:mmpf}
Z_N = \frac{1}{\mbox{vol}(U(N))} \int D M ~ e^{-\frac{1}{\la}
  \Tr~ W(M)}.   
\end{align}
In the large $N$ limit the distribution of the
eigenvalues $\la_i$ of $M$ on the real axis becomes continuous and defines a hyperelliptic curve. This curve is called the spectral curve of the matrix model.

In the 't Hooft limit this matrix model has a dual description as
the B-model topological string on Calabi-Yau geometries of the form \cite{dv-mm-ts}
\begin{align}\label{eq:DVgeometry}
uv + y^2 - W'(x)^2 + f(x) =0,
\end{align}
where $f(x) = 4 \mu \sum_{j=0}^{d-1} b_j x^j$ is a polynomial in $x$ of degree $d-1$. The hyperelliptic curve~$\Sigma$ modeling the local threefold equals the matrix model spectral curve, with 
\begin{align}
f(x) = \frac{4 \mu}{N} \sum_{i=1}^N
\frac{W'(x)-W'(\lambda_i)}{x-\lambda_i}. 
\end{align}
The potential $W(x)$ determines the positions of the cuts, containing
the non-normalizable moduli, while the size of the cuts is determined
by the polynomial $f(x)$, comprising the normalizable moduli $b_0,
\ldots, b_{d-2}$ and the log-normalizable modulus $b_{d-1}$.

This duality may be generalized by starting with multi-matrix models, whose spectral curve is a generic (in contrast to hyperelliptic) curve in the variables $x$ and $y$.   

The I-brane picture suggests that the full B-model partition function on these Calabi-Yau geometries can be understood in terms of $\cD$-modules. Even better, we will find that finite $N$ matrix models are determined by an underlying $\cD$-module structure. 

In the past, as well as recently, these matrix models have been studied in great detail in several contexts. Most importantly for us, it has been realized that a central role is played by the string or Douglas equation
\begin{align}
[P,Q] = \la.
\end{align}
Here, $P$ and $Q$ are operators that implement multiplication with and differentiation with respect to a spectral coordinate.   
In a double scaling limit $P$ and $Q$ turn into differential operators. Physically, these critical models are known to describe minimal string theory.  

Already in \cite{mooregeomstring,mooreisomo} an attempt has been made to understand this string equation in terms of a quantum curve in terms of the expansion in the parameter $\la$. In Moore's approach this surface seemed to emerge from an interpretation of the string equation as isomonodromy equations.

In topological as well as minimal string theory a dominant role is played by holomorphic branes: either topological B-branes \cite{adkmv} or FZZT branes  \cite{seibergshih,Maldacena:2004sn}. Their moduli space equals the spectral curve, whereas the branes themselves may be interpreted as fermions on the quantized spectral curve. 
In particular, for $(p,1)$ minimal models the so-called Lax operator $P$ has been interpreted as the quantization of the spectral curve. In these string theories
it is possible to compute correlation functions using a $W_{1+\infty}$-algebra \cite{adkmv,fukuma-winfty,fukuma-pq} that implements complex symplectomorphisms of the complex plane~$\mathcal{B}$ in quantum theory as Ward identities. 

These advances strongly hint at a fundamental appearance of $\cD$-modules in the theory of matrix models.
Indeed, this section unifies recent developments in matrix models in the framework of section~\ref{sec:formalism}. Firstly, after a self-contained introduction in double scaled models we uncover the $\cD$-module underlying the $(p,1)$-models. In the second part of this section we shift our focus to general Hermitian multi-matrix models, and unravel their $\cD$-module structure.

\subsection{Double scaled matrix models and the KdV hierarchy}

Our first goal is to find the $\cD$-modules that explain the quantum
structure of double scaled Hermitean matrix models. This double
scaling limit is a large $N$ limit in which one also fine-tunes the
parameters to find the right critical behaviour of the multi-matrix model
potential. Geometrically the double scaling limit zooms in on some
branch points  
of the spectral curve that move close together.
Spectral curves of double scaled matrix models are therefore of genus
zero and parametrized as 
\begin{align}
\Sigma_{p,q}: \quad y^p + x^{q} + \ldots =0.
\end{align} 
The one-matrix model only generates hyperelliptic spectral curves,
whereas the two-matrix model includes all possible combinations of $p$
and $q$.  
 These double scaled multi-matrix models are known
to describe non-critical ($c < 1$) bosonic string theory based on the
$(p,q)$ minimal model coupled to two-dimensional gravity
\cite{DVVloop,douglaspq,Daul:1993bg,
  review2dstuff,review2dgravity}. This field is 
therefore known as minimal 
string theory. 


Zooming in on a single branch point yields the geometry
\begin{align*}
\Sigma_{p,1}: \quad y^p = x, 
\end{align*}
corresponding to the $(p,1)$ topological minimal model. This model is strictly 
not a well-defined conformal field theory, but does make sense as
2d topological field theory. For $p=2$ it is known as topological
gravity \cite{wittentopphase, wittenintersectiontheory,
  kontsevichintersectiontheory, robbertcargese}. 
%
%
%
%

All $(p,q)$ minimal models turn out to be governed by two
differential operators 
\begin{align} 
P &= (\la \partial_x)^p + u_{p-2}(x) (\la \partial_x)^{p-2} + \ldots +
u_0(x), \\
Q &= (\la \partial_x)^q + v_{q-2}(x) (\la \partial_x)^{q-2} + \ldots +
v_0(x),
\end{align}
of degree $p$ and $q$ respectively, which obey the string (or Douglas)
equation 
\ba
[P,Q]= \la.
\end{align}
%
%
%
%
$P$ and $Q$ depend on an infinite set of times
$t= (t_1, t_2, t_3, \ldots )$, which are closed string
couplings in minimal string theory, and evolve in these times as  
\ba
\la \frac{\partial}{\partial t_j } P &= [(P^{j/p})_+,P],\\
\la \frac{\partial}{\partial t_j } Q &= [(P^{j/p})_+,Q],
\end{align} 
The fractional powers of
$P$ define a basis of commuting Hamiltonians\footnote{Notice that
  $L = P^{1/p}$ is a
  pseudo-differential operator, having  an expansion
\begin{align}
L = \la \partial_x + l_0(x) + l_1(x) (\la \partial_x)^{-1} + l_2(x)
(\la \partial_x)^{-2} + \ldots,
\end{align}
in negative powers of $\la \partial_x$. This extended notion of a
derivative is defined by the Leibnitz rule 
\begin{align}
\partial_x^n f = \sum_{k=0}^{\infty} { n \choose k} ( \partial_x^k f ) \partial^{n-k},
\end{align}
for any $n \in \Z$ with ${n \choose k} = n \cdot \ldots
\cdot (n-k+1) / k!$. It gives the derivatives with $n <0$ an
interpretation of partial integration. $L_+$ is the notation 
for the restriction to the positive powers of $L$. 
}.
This integrable system is known as the $p$-th KdV hierarchy and the
above evolution equations as the KdV flows.

The differential operator $Q$ is completely determined as a function of
fractional powers of the Lax operator $P$ and the times $t$
\ba
Q = - \sum_{{\tiny \begin{array}{c} j \ge 1 \\  j \neq 0 \mod
      p \end{array}}} \left( 1+ \frac{j}{p} \right) t_{j+p} P^{j/p}_+, 
\end{align}
This implies that when we turn off all the KdV times except for
$t_1=x$ and fix $t_{p+1}$ to be constant we find $Q =
\la \partial_x$. This defines the $(p,1)$-models  
\ba \label{eq:(p,1)-model}
P = (\la \partial_x)^p - x, \quad Q = \la \partial_x.
\end{align}
One can reach any other $(p,q)$ model by flowing in the
times $t$. 

The partition function of the $p$-th KdV hierarchy is a
tau-function as in equation~(\ref{taufunction}). The associated
subspace $\cW \in Gr$ may be found by studying the eigenfunctions
$\psi(t,z)$ of the Lax operator $P$
\begin{align}
P \psi(t,z) = z^p \psi(t,z).
\end{align}
The so-called Baker function $\psi(t,z)$ represents the fermionic field that
sweeps out the subspace $\cW$ in the times $t$. 

If we restrict to the $(p,1)$-models the Baker function $\psi(x,z)$
can be expanded in a Taylor series
\begin{align}
\psi(x,z) = \sum_{k=0}^{\infty} v_k(z) \frac{x^k}{k!}.
\end{align}
Since $\psi(x,z)$ is an element of $\cW$ for all times, this defines a
basis $\{ v_k(z)\}_{k \ge 0}$ of the subspace $\cW$. 
In fact, it is not hard to see that the $(p,1)$ Baker function is
given by the generalized Airy function
\begin{align}\label{eqn:Airyfunction}
\psi(x,z) =  e^{\frac{p z^{p+1}}{(p+1) \la}} \sqrt{z^{p-1}} \int dw
~e^{\frac{(-1)^{1/p+1} (x+z^p)w}{\la^{p/p+1}} +
  \frac{w^{p+1}}{p+1}}, 
\end{align}
which is normalized such that its Taylor components $v_k(z)$ can
be expanded as 
\begin{align}\label{eqn:(p,1)state}
v_k(z) = z^k (1 + \cO(\la/z^{p+1}))
\end{align}
The $(p,1)$ model thus determines the fermionic state
\begin{align}\label{eq:(p,1)state}
| \cW \rangle = v_0 \wedge v_1 \wedge v_2 \wedge \ldots,
\end{align}
where the $v_k(z)$ can be written explicitly in terms of Airy-like integrals
(see  \cite{robbertcargese} for a nice review). The invariance under
\begin{align}
z^p \cdot \cW \subset \cW 
\end{align}
characterizes this state as coming from a $p$-th KdV hierarchy. In the other
direction, the state $|\cW \rangle$ determines the Baker function (and
thus the Lax operator) as
the one-point function
\begin{align}
\psi(t,z) = \langle t | \psi(z) |\cW \rangle.
\end{align}

In the dispersionless limit $\la \to 0$ the derivative
$\la \partial_x$ is replaced by a variable $d$, and the Dirac
commutators by Poisson brackets in $x$ and $d$. The leading order
contribution to the string equation is given by the Poisson bracket
\begin{align}
\{ P_0 , Q_0 \} =1,
\end{align}
where $P_0$ and $Q_0$ equal $P$ and $Q$ at $\la =0$. The solution to
this equation is 
\begin{align}
P_0(d;t) &= x \\
Q_0(d;t) &= y(x;t) 
\end{align}
and recovers the genus zero spectral curve $\Sigma_{p,q}$ of the double
scaled matrix model, parametrized by $d$. The KdV flows \emph{deform} this
surface in such a way that its singularities are preserved. (See the
appendix of \cite{Maldacena:2004sn} for a detailed discussion.) 

Note that $\Sigma_{p,q}$ is not a spectral curve for the
Krichever map. The Krichever curve is instead found as the
space of simultaneous eigenvalues of the differential operators 
\begin{align}
[P,Q]=0,
\end{align}
that is \emph{preserved} by the KdV flow as a straight-line flow along its
Jacobian. In fact, there is no such Krichever spectral curve corresponding to
the doubled scaled matrix model solutions. 

Wrapping an I-brane around $\Sigma_{p,q}$
quantizes the semi-classical fermions on the spectral curve
$\Sigma_{p,q}$. 
The only point at infinity on $\Sigma_{p,q}$ is given by $x \to \infty$. 
The KdV tau-function should thus be the
fermionic determinant of the quantum state $| \cW \rangle$
that corresponds to this $\cD$-module. In the next subsection we write
down the 
$\cD$-module describing the $(p,1)$ model and show precisely how
this reproduces the tau-function using the prescription outlined in
section~\ref{sec:formalism}.

\subsection{$\cD$-module for topological
  gravity}\label{sec:minimalmodels}

We are ready to reconstruct the $\cD$-module that yields the
fermionic state $|\cW \rangle$ in equation~(\ref{eq:(p,1)state}).  
For simplicity we study the $(2,1)$-model, associated to an I-brane
wrapping the curve  
\begin{align}
\Sigma_{(2,1)}: \qquad y^2 = x \qquad (\mbox{with}~x,y \in \C).
\end{align}
Notice that this is an $2:1$ cover over the $x$-plane. It contains
just one asymptotic region, where $x \to \infty$.
Fermions on this 
cover will therefore sweep out a subspace $\cW$ in the Hilbert space
\begin{align}
\cW  ~\subset~ \cH(S^1) = \C((y^{-1})), 
\end{align}
the space of
formal Laurent series in $y^{-1}$. The fermionic vacuum $|0 \rangle \subset \cH(S^1)$
corresponds to the subspace 
\begin{align}
|0 \rangle =  y^{1/2} \wedge y^{3/2} \wedge y^{5/2} \wedge \ldots,
\end{align}
which encodes the algebra of functions on the disk parametrized by $y$
and with boundary at $y= \infty$. Exponentials in $y^{-1}$
represent non-trivial behaviour near the origin and therefore act
non-trivially on the vacuum state. In contrast, exponentials in $y$ are 
holomorphic on the disk and thus act trivially on the vacuum. 

The $B$-field $B = \frac{1}{\la} dx \wedge dy$ quantizes the algebra of
functions on $\C^2$ into the differential algebra 
\begin{align}
\cD_{\la} = \langle x, \la \partial_x \rangle.
\end{align}
Furthermore, it introduces a meromorphic connection 1-form
$A  = \frac{1}{\la} y dx$
on $\Sigma_{(2,1)}$, which pushes forward to the rank two $\la$-connection 
\begin{align}\label{eq:(2,1)connection}
\nabla_A = \la \partial_x-  \left( \begin{array}{cc} 0
    & 1 \\ x & 0  \end{array} \right) 
\end{align}
on the base, parametrized by $x$. We claim that the
corresponding $\cD_{\la}$-module $\cM$, generated by  
%
 %
\bea
P= (\la \partial_x)^2 -x, 
\eea
describes the $(2,1)$ model. Let us
verify this. 

Trivializing the $\la$-connection $\nabla_A$ in (\ref{eq:(2,1)connection}) implies
finding a rank two matrix $g(x)$ such that
\begin{align*}
\nabla_A =  \la \partial_x - g'(x) \circ g^{-1}(x).
\end{align*}
The columns of $g$ define a basis of solutions $\Psi(x)$ to the
differential equation $\nabla_A \Psi(x) = 0.$
They are meromorphic flat sections for
$\nabla_A$ that determine a trivialization of the bundle near $x =
\infty$. As the connection $\nabla_A$ is pushed forward from the
cover, $\Psi(x)$ is of the form
\begin{align*}
\Psi(x) = \left( \begin{array}{c} \psi(x) \\ 
    \psi'(x) \end{array} \right). 
\end{align*}

Independent solutions have different asymptotics in the semi-classical
regime where $x \to \infty$. In the $(2,1)$-model the two independent solutions $\psi_{\pm}(x)$ solve the differential equation 
\begin{align}
P \psi_{\pm}(x) = ((\la \partial_x)^2 - x) \psi_{\pm}(x) =0.
\end{align}
Hence these are the functions $\psi_+(x) =$ Ai$(x)$ and $\psi_-(x) =$
Bi$(x)$, that correspond semi-classically to the two saddles 
\begin{align*}
w_{\pm} = \pm \sqrt{x} / \la^{1/3}
\end{align*}
of the Airy integral  
\begin{align}
\psi(x) = \frac{1}{2 \pi i} \int dw~e^{-\frac{xw}{\la^{2/3}} +
  \frac{w^3}{3}}.
\end{align}

The $\cD$-module $\cM$  can be quantized
into a fermionic state for any choice of boundary
conditions. Depending on this choice we find an  $\cO(x)$-module $\cW_{\pm}$
spanned by linear combinations of $\psi_{\pm}(x)$ and of $\psi_{\pm}'(x)$.  
The fermionic state is generated by asymptotic
expansions in the parameter $\la$ of these elements.

The saddle-point approximation around the saddle $w_{\pm} = \pm
\sqrt{x}/\la^{1/3}$ yields
\begin{align*}
\hspace*{1.3cm} \psi_{\pm}(x) & \sim   y^{-1/2}~ e^{\mp \frac{2  
    y^{3}}{3 \la}} \left( 1 + \sum_{n \ge 1} c_n \la^n (\pm y)^{-3n}
\right) \notag \\
&\sim   y^{-1/2}~ e^{\mp \frac{2 
    y^{3}}{3 \la}}~ v_0(\pm y).
\end{align*} 
To see the last step just recall the definition of
$v_0(z)$ as being equal to the Baker function $\psi(x,z)$ evaluated at
$x=0$.\footnote{Remark that $x$ and $z^2$ appear equivalently in
$\psi(x,z)$, while $\psi(x)$ and $\psi(x,z)$ only differ in the
normalization term in $z$.}
A similar expansion can be made for $ \psi'(x)$ with the result 
\begin{align*}
\hspace*{-1.3cm} \psi_{\pm}'(x) &\sim y^{1/2}~
e^{\mp \frac{ 2  y^{3}}{3 \la}}~ v_1(\pm y).
\end{align*} 

Note that both expansions in $\lambda$ are functions in
the coordinate $y$ on the cover. They contain a classical term (the
exponential in $1/\la$), a 1-loop piece and a quantum expansion in $\la
y^{-3}$. When we restrict to the saddle $w = \sqrt{x}/\la^{1/3}$,
these series blend the into the fermionic state
\begin{align}
| \cW_+ \rangle = \psi_+(y) \wedge \psi_+'(y) \wedge y^p \psi_+(y)
  \wedge y^p \psi_+'(y) \wedge \ldots. 
\end{align} 
Does this agree with the well-known result~(\ref{eq:(p,1)state})?  

First of all, notice that the basis vectors $x^k \psi(x)$ and $x^k
\psi'(x)$, with $k>0$, contain in their expansions the function
$v_k(y)$ plus a sum 
of lower order terms in $v_l(y)$ (with $l <k$). The wedge
product obviously eliminates all these lower order terms. Secondly, the
extra factor $y^{-1/2}$ factors just reminds us that we have written 
down a fermionic state. 

Furthermore, 
the WKB exponentials are exponentials in $y$ and thus elements of
the subgroup $\Gamma_+$ of holomorphic functions that extend over a disc centered around $y=0$, whereas the expansions $v_k(y)/y^k$ are part of the subgroup $\Gamma_-$ of functions that extend over a disc centered at $y = \infty$. (The definition of these subgroups and their action on the infinite Grassmannian can be found in Appendix~\ref{sec:infgrassmannian}.) 
Up to normal ordening ambiguities this shows that the WKB part gives a
trivial contribution to the fermionic state $|\cW_+ \rangle$. In fact, the tau-function even 
cancels these ambiguities. 

This shows that 
\begin{align}
| \cW_+ \rangle = v_0(y) \wedge v_1(y) \wedge v_2(y) \wedge \ldots
\end{align} 
is indeed the same as in (\ref{eq:(p,1)state}), when we change
variables from $z$ to $y$ in that equation. Of course, this doesn't change the
tau-function. 

So our conclusion is that the $\cD$-module underlying
topological gravity is the canonical $\cD$-module 
\begin{align}
\cM = \frac{\cD_{\la}}{\cD_{\la}((\la \partial_x)^2 -x)}. 
\end{align}
This $\cD$-module gives the definition of the quantum curve
corresponding to the $(2,1)$ model and defines its quantum partition
function in an expansion around $\la=0$. Exactly the same reasoning
holds for the $(p,1)$-model, where we find 
a canonical rank $p$ connection on the base.  It would be nice to be
able to write down a $\cD$-module for general $(p,q)$-models as
well.

\subsection{$\cD$-module for Hermitean matrix models}

$\cD$-modules continue to play an important role in
any Hermitean matrix model. In this subsection we are guided by 
\cite{eynard-isomonodromic} and \cite{BEH-duality,2-matrixdmodule} of Bertola,
Eynard and Harnad. 

We first summarize 
how the partition function for a 1-matrix model defines a tau-function for
the KP hierarchy. As we saw before, such a tau-function corresponds
to a fermionic state $|\cW \rangle$, whose basis elements we will write
down. Following \cite{eynard-isomonodromic} we discover a rank two
differential structure in this basis, whose determinant reduces to the
spectral curve in the semi-classical limit. This $\cD$-module
structure is somewhat more complicated then the $\cD$-module we just
found describing double scaled matrix models.

We continue with 2-matrix models, based on
\cite{2-matrixdmodule}. Instead of one differential equation, these
models determine a group of four differential equations,
that characterize the $\cD$-module in the local coordinates $z$ and
$w$ at infinity. The matrix model partition function may of course be
computed in either frame.

\subsub{1-matrix model}

Let us start with the 1-matrix model partition function
(\ref{eq:mmpf}). By diagonalizing the matrix $M$ the matrix integral
may be reduced to an integral over the eigenvalues $\la_i$ 
\begin{align}
Z_N = \int \prod_i d \la_i ~ \Delta(\la)^2 ~ e^{- \frac{1}{\la} \sum_i
  W(\la_i)},  
\end{align}
with the Vandermonde determinant $\Delta(\la) = \prod_{i<j} (\la_i -
\la_j) = \det(\la^{j-1}_i)$. The method of orthogonal polynomials
solves this integral by introducing an infinite set of polynomials $p_k(x)$, defined by the properties
\begin{align}
& \quad p_k(x) = x^k (1 + \cO(x^{-1})),\\
& \int dx ~ p_k(x)~ p_l(x)~ e^{-\frac{1}{\la} W(x)} = h_k \delta_{k,l}.
\end{align} 
The normalization of their leading term determines the coefficients
$h_n \in \C$. Since the Vandermonde determinant $\Delta(x)$ is not
sensitive to exchanging its entries $x^{j-1}_i$ for $p_{j-1}(x_i)$,
substituting $\Delta(x) = \det(p_{j-1}(x_i))$ turns the partition
function into a product of coefficients
\begin{align}
Z_N = N! \prod_{k=0}^{N-1} h_k.   
\end{align}
With the help of orthogonal polynomials the large $N$ behaviour of
$Z_N$ may be studied, while keeping track of $1/N$ corrections.

In this discussion the orthogonal polynomials are relevant since they
build up a basis for 
the fermionic KP state. In an appendix of \cite{eynard-isomonodromic}
it is shown that one should start at $t=0$  with a state $|\cW_{0}
\rangle$ generated by the polynomials $p_k(x)$ for $k \ge N$
\begin{align}
|\cW_{0} \rangle = p_N(x) \wedge p_{N+1}(x) \wedge p_{N+2}(x) \wedge \ldots.
\end{align}
Notice that the vector $p_{N}(x)$ thus corresponds to
the Fermi level and defines the Baker function in the double scaling
limit. Acting on them with the commuting flow generated by  
\begin{align}
\Gamma_{+} = \left\{ g(t) = e^{\sum_{n \ge 1} \frac{1}{n} t_n x^n} \right\}   
\end{align}
defines a state  $|\cW_{t} \rangle = | g(t) \cW_{0} \rangle$ at time
$t$, which allows to compute a tau-function at time $t$. If the
coefficients $u_j$ in the potential $W(x)$ are taken to be $u_j =
u_j^{(0)} + t_j$, this $\tau$-function equals the ratio of the matrix
model partition function $Z_N$ at time $t$ divided by that at
$t=0$.

Multiplying the orthogonal polyonomials by $\exp(-\frac{1}{2
  \la}W(x))$ doesn't change the fermionic state $\cW = \cW_0$ in a
relevant way, since this
factor is an 
element of $\Gamma_{+}$. To find the right $\cD$-module structure, it is
necessary to proceed with the quasi-polynomials 
\begin{align}
\psi_k(x) = \frac{1}{\sqrt{h_k}} p_k e^{- \frac{1}{2\la} W(x)}, 
\end{align}
which form an orthonormal basis with respect to the bilinear form
\begin{align}\label{eq:mmbilinearform}
(\psi_k, \psi_l) = \int dx~\psi_k \psi_l .
\end{align}
 
It is possible to express both multiplication by $x$ and
differentiation with respect to $x$ in terms of the basis of
$\psi_m$'s. The Weyl algebra $\langle x, \la \partial_x \rangle$ acts on
these (quasi)-polynomials by two matrices $Q$ and $P$
\begin{align}
x \psi_k(x) &= \sum_{l=0}^{\infty} Q_{kl} \psi_l \\
\la \partial_x \psi_k(x) &= \sum_{l=0}^{\infty} P_{kl} \psi_l(x),
 \end{align}
and the space of quasi-polynomials $\psi_k$ is thus a
$\cD_{\la}$-module.\\

Notice that we anticipate that the $\cD$-module possesses a rank two structure, since we started with a flat connection $A
=\frac{1}{\la} ydx$ on 
an I-brane wrapped on a hyper-elliptic curve. Now, the matrices $Q$
and $P$ only contain non-zero entries in a finite band around the
diagonal. The action
of $\partial_x$ on the semi-infinite set of $\psi_k(x)$'s can
therefore indeed be summarized in a rank two  
differential system (\cite{eynard-isomonodromic} and references
therein)   
\begin{align}\label{eq:MMDmodule}
\la \partial_x \left[ \begin{array}{c} \psi_{N}(x) \\
    \psi_{N-1}(x) \end{array} \right] = A_N(x)
\left[ \begin{array}{c} \psi_{N}(x) \\ \psi_{N-1}(x) \end{array}
\right], 
\end{align}
where $A_N(x)$ is a rather complicated $2 \times
2$-matrix involving 
the derivative $W'$ of the potential and the infinite matrix $Q$:
\begin{align}
A_N(x) =  \frac{1}{2} W'(x) \left[ \begin{array}{cc} -1 & 0 \\ 0 & 1 \end{array} \right] &
+ \gamma_N  
\left[ \begin{array}{ll} -\widetilde{W}'(Q,x)_{N,N-1} &
    \widetilde{W}'(Q,x)_{N,N}\\  -\widetilde{W}'(Q,x)_{N-1,N-1} &
    \widetilde{W}'(Q,x)_{N-1,N} \end{array} \right], 
\end{align}
with
\begin{align} 
\widetilde{W}'(Q,x) = \left( \frac{W'(Q)-W'(x)}{Q-x} \right)\quad
\mbox{and} \quad 
\gamma_N = \sqrt{\frac{h_N}{h_{N-1}}}.
\end{align}
Equation (\ref{eq:MMDmodule}) is thus the rank two $\la$-connection
defining the $\cD_{\la}$-module structure on $\cW$ that we were
searching for!  As a check, the determinant of this connection reduces
to the spectral 
curve in the semiclassical, or dispersionless, limit
\cite{eynard-isomonodromic}:  
\begin{align}
  \Sigma_N: \quad 0 &= \mbox{det} \left( y 1_{2 \times 2} - A_N(x)
  \right) \\& = y^2 - W'(x)^2 + 4 \la \sum_{j=0}^{N-1} \left( \frac{W'(Q) - W'(x)}{Q-x} \right)_{jj}
\end{align}
(To make the coefficients in the above equation agree with
(\ref{eq:DVgeometry}), we rescaled $y \mapsto y/2$.)
In conclusion we found the $\cD$-module structure underlying Hermitean
1-matrix models. 

Remark that in the $N \to \infty$ limit we expect that the hyperelliptic
curve defining the B-model Calabi-Yau (\ref{eq:DVgeometry}) emerges
from $\Sigma_N$. Indeed, in the 't Hooft limit $Q$
corresponds classically to the
coordinate $x$ on the curve, whereas quantum-mechanically it is an
operator whose spectrum is described by the eigenvalues $\la_i$ of
the infinite matrix $M$. In the large $N$ limit we can therefore
replace the matrix $Q_{ij}$ in the definition for $\Sigma_N$ by
$\lambda_i \delta_{ij}$. 

Note as well that we can rewrite the rank two connection for
the vector $(\psi_N, \psi_N')^t$ as 
\begin{align}
\la \partial_x \left[ \begin{array}{c} \psi_{N}(x) \\
    \psi'_{N}(x) \end{array} \right] = \left[ \begin{array}{ll}
   0 & 
    1 \\  -\det(A_N(x)) + \la Y  &
    \la Z  \end{array} \right]
\left[ \begin{array}{c} \psi_{N}(x) \\ \psi'_{N}(x) \end{array}
\right],
\end{align}
at least when tr$(A_N(x))=0$,
with $Y$ and $Z$ some derivatives of entries of $A_N(x)$.
This brings the $\la$-connection in the familiar form
of section~\ref{sec:formalism}. In the next subsection we clarify the
differential structure in a simple example.
  
\subsub{2-matrix model}

Let us first say a few words on the $\cD$-module structure underlying
multi-matrix models, which capture spectral curves of any degree in
$x$ and $y$ \cite{BEH-duality, 2-matrixdmodule}. The partition function
for a two-matrix model, with two rank $N$ matrices $M_1$ and $M_2$, is 
\begin{align}
Z_N =  \int DM_1 DM_2 ~e^{- \frac{1}{\la}  \Tr \left( W_1(M_1) +
    W_2(M_2) - M_1 M_2 
  \right) }, 
\end{align} 
where $W_1$ and $W_2$ are two potentials of degree $d_1+1$ and
$d_2+1$. Choosing $W_2$ to be
Gaussian reduces the 2-matrix model to a 1-matrix model. The 2-matrix
model is solved by introducing two sets of orthogonal polynomials
$\pi_k(x)$ 
and $\sigma_k(y)$. Again it is convenient to turn them into quasi-polynomials 
\begin{align}
\psi_k(x) =  \pi_k(x) e^{- \frac{1}{\la} W_1(x)} , \quad
\phi_k(y) = \sigma_k(y) e^{- \frac{1}{\la} W_2(y)}.
\end{align}
obeying the orthogonality relations 
\begin{align}\label{eq:2-matrixortho}
\int dx dy ~\psi_k(x) \phi_l(y) e^{\frac{xy}{\la}} = h_k \delta_{kl}. 
\end{align}

Multiplying with or taking a derivative with respect to either $x$ or
$y$ yields (just) two operators $Q$ and $P$ (and their transposes
because of (\ref{eq:2-matrixortho})), that form a representation of
string equation $[P,Q]=0$. Since $Q$ is only non-zero in a band around
the diagonal of size $d_2+1$ and $P$ of size $d_1+1$, the
quasi-polynomials may be folded into the vectors 
\begin{align}
\vec{\psi} = [ \psi_{N}, \ldots, \psi_{N-d_2} ]^t, \quad
\vec{\phi} = [ \phi_{N}, \ldots, \phi_{N-d_1} ]^t.
\end{align}
Any other quasi-polynomial can be expressed as a sum of entrees of
these vectors, with coefficients in the polynomials in $x$ and $y$. 
These vectors are called windows. The differential operators
$\la \partial_x$ 
and $\la \partial_y$ respect them, so that their action is
summarized in a rank $d_2+1$ resp. rank $d_1+1$ $\la$-connection
\begin{align}\label{eq:2matrixconnection}
\la \partial_x \vec{\psi}(x) = A_1(x) \vec{\psi}(x) , \quad
\la \partial_y \vec{\phi}(y) =  A_2(y) \vec{\phi}(x).
\end{align}
This we interpret as two representations of the $\cD_{\la}$-module underlying 2-matrix
models. Indeed, \cite{BEH-duality} proves that the determinant of both
differential systems equals the same spectral curve $\Sigma$, in the
limit $\la \to 0$ when we 
replace $\la \partial_x \to y$ and  $\la \partial_y \to x$. The
defining equation of $\Sigma$ is of degree $d_1+1$ in $x$ and of
degree $d_2+1$ in $y$.  

In fact, it is useful to introduce two more semi-infinite sets of
quasi-polynomials $\underline{\psi}_k(y)$ and $\underline{\phi}_k(x)$,
as the Fourier transforms of $\psi_k(x)$ and $\phi_k(y)$
respectively. The action of the Weyl algebra on them may be encoded as
the transpose of the above linear systems. The full system can
therefore be summarized by (compare to (\ref{eq:c=1module}))
\begin{align}
x\mbox{-axis}: \quad \{ \psi_k(x),~\underline{\phi}_k(x)\}, \quad
& \nabla_{\la} = \la \partial_x - A_1(x), \\
y\mbox{-axis}: \quad \{\phi_k(y),~\underline{\psi}_k(y)\}, \quad &
\nabla_{\la} = \la \partial_y - A_2(y). \notag
\end{align}
Moreover, the matrix model partition function can be rewritten as a
fermionic correlator in either local coordinate
\begin{align}
Z_N &= \int \prod_i d \la_i^{1} d \la_i^2~ \Delta(\la^1) \Delta(\la^2)~
e^{- \frac{1}{\la} \sum_i ( W_1(\la^1_i) + W_2(\la^2_i) - \la^1_i
  \la^2_i} \\
&= N! \prod_{k=0}^{N-1} \langle \psi_k(x)  |  \underline{\phi}_k(x) \rangle 
= N! \prod_{k=0}^{N-1} \langle \phi_k(y)  |  \underline{\psi}_k(y)
\rangle \notag 
\end{align}
with respect to the bilinear form in (\ref{eq:mmbilinearform}).

Furthermore, Bertola, Eynard and Harnad study the dependence on the
parameters $u^{(1)}_j$ and $u^{(2)}_j$ appearing in the potentials
$W_1$ and $W_2$. Deformations in these parameters leave the
two sets of quasi-polynomials invariant as well. On
$\vec{\psi}$ and $\vec{\phi}$  they act as matrices $U^{(1)}_j$ and
$U^{(2)}_j$. This yields the 2-Toda system 
\begin{align}
\partial_{u^{(1)}_j} Q = - [Q,U^{(1)}_j] \qquad \partial_{u^{(1)}_j}
P = - [P,U^{(1)}_j] \\
\partial_{u^{(2)}_j} Q =  [Q,U^{(2)}_j] \qquad \partial_{u^{(2)}_j},
P = [P,U^{(1)}_j]. 
\end{align}
In \cite{BEH-duality} it is proved that the linear differential systems
(\ref{eq:2matrixconnection}) are compatible with these
deformations, so that the parameters $u^{(1)}_j$ and $u^{(2)}_j$
in fact generate isomonodromic deformations. This shows precisely how the
non-normalizable parameters in the potential respect the
central role of the $\cD_{\la}$-module (\ref{eq:2matrixconnection}) in the 
2-matrix model.

\subsection{Gaussian example}

Let us consider the Gaussian matrix model with quadratic potential 
\begin{align}
 W(x) =  \frac{x^2}{2}, 
\end{align}
that is associated to the spectral curve
\begin{align}\label{eq:spectralcurvegaussian}
y^2 = x^2 - 4 \mu^2
\end{align}
in the large $N$ limit.  In the Dijkgraaf-Vafa correspondence this
matrix model is thus dual to the topological 
B-model on the deformed conifold geometry.

The Hermite functions 
\begin{align*}
 \psi^{\la}_k(x) &=  \frac{1}{\sqrt{h_k}} e^{-\frac{x^2}{4 \la}}
 H^{\la}_k(x), \quad \mbox{with}\\ 
 H^{\la}_k(x)&=  \la^{k/2} H_k\left( \frac{x}{\sqrt{\la}} \right) =
 x^k \left( 1 + \cO\left( \frac{\sqrt{\la}}{x} \right) \right),
\end{align*}
form an orthogonal basis for this model. Their inner product is given
by
\begin{align*}
\int \frac{dx}{2 \pi} ~ \psi^{\la}_k(x)  \psi^{\la}_l(x) =
 \la^k k! \sqrt{\frac{\la}{2 \pi}} \delta_{kl} \quad \Rightarrow \quad
 h_k =    \la^k k! \sqrt{\frac{\la}{2 \pi}}.  
\end{align*}
The partition function of the Gaussian matrix model can be computed as
a product of the normalization constants $h_k$. Using the asymptotic
expansion of the Barnes function $G_2(z)$, that is defined by
$G_2(z+1) = \Gamma(z) G_2(z)$, the free energy can be expanded in powers 
of $\la$ 
\begin{align}\label{eqn:Gaussianpartfunc}
 \cF_N &=  \log \prod_{k=1}^{N-1} h_k = \log \left(  G_2(N+1)
   \frac{\la^{N^2/2}}{(2  \pi)^{N/2}} \right) \\
& = \frac{1}{2} \left( \frac{\mu}{\la} \right)^2 \left( \log  \mu - \frac{3}{2} \right) -
\frac{1}{12} \log \mu + \zeta'(-1) +
\sum_{g=2}^{\infty} 
\frac{B_{2g}}{2g(2g-2)} \left( \frac{\la}{\mu} \right)^{2g-2} \notag,
\end{align}
where $B_{2g}$ are the Bernoulli numbers and $\mu = N \la$.  

The derivatives of the Hermite functions are related as 
\begin{align}
\la \frac{d}{dx} \left[ \begin{array}{cc} \psi^{\la}_{k}(x) \\
    \psi^{\la}_{k-1}(x) \end{array} \right] = 
\left[ \begin{array}{cc} -x/2 & \sqrt{k \la} \\ -\sqrt{k \la} &
    x/2\end{array} \right] \left[ \begin{array}{cc} \psi^{\la}_{k}(x) \\
    \psi^{\la}_{k-1}(x) \end{array} \right]. 
\end{align}
So, according to the previous discussion, the $\cD_{\la}$-module
connection is given by 
\begin{align}\label{eq:Dmodulegaussian}
\la \frac{d}{dx} - A_N(x) = \la \frac{d}{dx} +
\left[ \begin{array}{cc} x/2 & -\sqrt{N \la} \\ \sqrt{N \la} &
   - x/2 \end{array} \right]. 
\end{align}
Here we choose $\vec{\psi} = [ \psi_N, \psi_{N-1}]^t$ as window.
In the large $N$ limit the determinant of this rank two
differential system indeed yields the spectral curve
(\ref{eq:spectralcurvegaussian}) with $\mu = \la N$.  

Instead of using $\psi_{k}^{\la}$ and $\psi_{k-1}^{\la}$ as a basis,
we can also write down the differential system for $\psi_{k}^{\la}$
and its derivative ${\psi'}_{k}^{\la}(x)= \la \partial_x
\psi_{k}^{\la}(x)$. Since this 
derivative is a linear combination of $\psi_{k-1}^{\la}$ and $x
\psi_{k}(x)$ (as we saw above), it is equivalent to use this basis to generate
the fermionic state $\cW $. We compute that 
\begin{align}
\la \frac{d}{dx} \left[ \begin{array}{cc} \psi^{\la}_{N}(x) \\
    {\psi'}^{\la}_{N}(x) \end{array} \right] = 
\left[ \begin{array}{cc} 0 & 1 \\ x^2 - \la N - \la/2 &
    0 \end{array} \right] \left[ \begin{array}{cc} \psi^{\la}_{N}(x) \\
    {\psi'}^{\la}_{N}(x) \end{array} \right]. 
\end{align}
The spectral curve in the large $N$ limit hasn't changed. Notice that
in this form it is clear that the rank 2 connection is the
push-forward of the connection $A = \frac{1}{\la} y dx$ on the
spectral curve $y^2 = x^2 - 4 \mu$ to the $\C$-plane, up to some $\la$-corrections.

In the double scaling limit the limits $N \to \infty$ and $\la \to 0$
are not independent as in the 't Hooft limit, but correlated, such
that the higher genus contributions to the partition function are
taken into account. In terms of the Gaussian spectral curve this limit implies
that one zooms in onto one of the endpoints of the cuts. The
orthogonal function $\psi^{\la}_N(x)$ turns into the Baker function
$\psi(x)$ of the double scaled state $\cW$.
 
In the Gaussian matrix model this is implemented by letting $x \to 
\sqrt{\mu} + \epsilon x$, where $\epsilon$ is a small parameter. So
the double scaled spectral curve reads  
\begin{align}
y^2 = x,
\end{align}
while the differential system reduces to 
\begin{align}
\la \frac{d}{dx} \left[ \begin{array}{cc} \psi(x) \\
    \psi'(x) \end{array} \right] = 
\left[ \begin{array}{cc} 0 & 1 \\ x &
    0 \end{array} \right] \left[ \begin{array}{cc} \psi(x) \\
    \psi'(x) \end{array} \right].
\end{align}
This is indeed the $\cD$-module corresponding to the $(2,1)$-model.

\section{Conifold and $c=1$ string}\label{sec:c=1}

The free energy~(\ref{eqn:Gaussianpartfunc}) of the Gaussian matrix
model pops up in the theory of bosonic $c=1$ strings. This $c=1$
string theory 
is formulated in terms of a single bosonic coordinate $X$, that is
compactified on a circle of radius $r$ in the Euclidean theory. A
critical bosonic string theory (with 
$c = 26$) is obtained by coupling the above CFT to a Liouville
field $\phi$. The Liouville field corresponds to the non-decoupled conformal mode of
the worldsheet metric. The local worldsheet action reads 
\begin{align*}
\frac{1}{4 \pi} \int d^2 \sigma \left(\frac{1}{2} (\d X)^2 +  (\d \phi)^2
   +  \mu e^{\sqrt{2} \phi} + \sqrt{2} \phi R  \right),
\end{align*}
where the coupling $\mu$ is seen as the worldsheet cosmological
constant. In the Euclidean model there are only two sets of operators, that
describe the winding and momenta modes of the field $X$. These vertex
and vortex operators can be added to the action as marginal
deformations with coefficients $t_n$ and $\tilde{t}_n$.  

Just like in $c<1$ minimal string theories (the
$(p,q)$-models of last section), the partition 
function of the $c=1$ string is first computed using a dual matrix
model description \cite{Gross:1990ub}.
At the self-dual radius $r=1$ it agrees with the Gaussian matrix
model partition function in equation~(\ref{eqn:Gaussianpartfunc}),
where $\la$ now plays the role of the $c=1$ string coupling constant.  

The matrix model dual to the $c=1$ string is called matrix quantum
mechanics. This duality is reviewed in much detail in \emph{e.g.}
\cite{Klebanov:1991qa,Polchinski:1994mb,Alexandrov:2003ut}. 
Matrix quantum mechanics is described by a gauge
field $A$ and a scalar 
field $M$ that are both $N \times N$ Hermitean matrices. The momentum
modes of the $c=1$ string correspond to excitations of $M$, whereas
the winding modes are excitations of $A$. If we focus on the momentum
modes, the (double scaled) matrix model is governed by the Hamiltonian
\begin{align*}
H = \frac{1}{2} \, \Tr \left( - \la^2 \frac{\d^2}{\d M^2} - M^2 \right).
\end{align*}

Let us focus on solutions that depend purely on the eigenvalues
$\la_i$ of $M$. The Hamiltonian may be rewritten in terms of the
eigenvalues as 
\begin{align*}
H = \frac{1}{2} \,  \Delta^{-1}(\la) \sum_i \left( - \la^2
    \frac{\d^2}{\d \la_i^2} - \la_i^2 \right) \Delta(\la),
\end{align*}
where $\Delta(\la)$ is Vandermonde determinant. It is convenient to
absorb the factor $\Delta$ in the wavefunction solutions, making them
anti-symmetric. Hence, the singlet sector of matrix quantum mechanics
describes a system of $N$ free fermions in an upside-down Gaussian
potential.   

To describe the partition function of the $c=1$ model it is convenient to move over to light-cone
coordinates $\la_{\pm} = \la \pm p$,
so that elementary excitations of the $c=1$ model are
represented as collective excitations of free fermions near the Fermi
level 
\begin{align}\label{eqn:fermilevel}
 \la_+ \la_- = \mu.
\end{align}
When we restrict to $\la_{\pm}>0$, scattering amplitudes can be
computed by preparing asymptotic free fermionic states $\langle \tilde{t} |$
and $| t \rangle$ at the regions where
one of $\la_{\pm}$ becomes very large.

In this picture the generating function of scattering amplitudes has a
particularly simple form. It can be formulated as a fermionic
correlator \cite{Dijkgraaf:1992hk}  
\begin{align}\label{c=1:toda}
Z =  \langle t |  S  | \tilde{t} \rangle, 
\end{align}
where the fermionic scattering matrix $S \in GL(\infty,\C)$ was first
computed in \cite{Moore:1991zv}.  Moreover, in
\cite{Alexandrov:2002fh} (see also Chapter V of 
\cite{Alexandrov:2003ut}) and later in \cite{adkmv} it is noticed that $S$ just equals the Fourier
transformation
\begin{align}
(S \psi) (\la_-) = \int d\la_+ \, e^{ \frac{1}{\la} \la_- \la_+} \psi(\la_+).\label{eqn:c=1fourier} 
\end{align}
In the next section we show that
this follows naturally from the perspective of $\cD$-modules. 

The result (\ref{c=1:toda}) shows that $c=1$ string theory is an integrable
system, just like the 
$(p,q)$-models in the last section. Since it depends on two
sets of times this integrable system is not a KP system. Instead, the
above expression defines a tau function of a 2-Toda
hierarchy. \index{Toda hierarchy}

%
%

Notice that the Fermi level (\ref{eqn:fermilevel}) is a real cycle on
the complex curve  
\begin{align}\label{conifold}
\Sigma: \quad  zw = \mu,
\end{align}
which is a different parametrization of the spectral curve $y^2 = x^2 -
\mu$ of the Gaussian 1-matrix model. 
In the revival of this subject a few years ago, a number of other
matrix model interpretations have been found. This includes a duality with the
Hermitean 2-matrix model, which makes the 2-Toda structure manifest
\cite{Bonora:1994ka}, a Kontsevich-type 
model \cite{Distler:1990mt,Imbimbo:1995yv} at the self-dual radius,
and a so-called normal matrix model 
\cite{Alexandrov:2003qk,Mukherjee:2005aq}, that parametrizes the dual
real cycle on the complex curve $\Sigma$. Let 
us also mention that the well-known duality of the $c=1$ string with
the topological B-model on the deformed conifold
\cite{Ghoshal:1995wm}, that follows, with a detour, from the more
general Dijkgraaf-Vafa correspondence.



\subsection{$\cD$-module description of the $c=1$ string}

This paragraph reproduces the $c=1$ partition
function (\ref{c=1:toda}) from a
$\cD$-module point of view. The discussion
continues the line of thought in section 5.5 of \cite{adkmv} and in \cite{dhsv}. 

As we have just seen, the $c=1$ string is geometrically characterized
by the presence of a holomorphic curve in $\C \times \C$ defined by 
\begin{align*}
\Sigma_{c=1}: \quad zw = \mu.
\end{align*}
Let us consider an I-brane wrapping the curve $\Sigma_{c=1}$. 
When we assume $z$ as local coordinate the curve quantizes into the
differential operator  
\begin{align}
P = - \la z\d_z - \mu.
\end{align}
It is amusing that the differential operator $P$ appears as a canonical
example in the theory of $\cD$-modules (see {\it e.g.}
\cite{kashiwara}) in the same way as the $c=1$ string is an elementary
example of a string theory. 

We recognize this example from section~\ref{sec:formalism}, where a
$\cD$-module was associated to the differential operator $P$. However, now it is important not to
forget that there are \emph{two} asymptotic points $z_{\infty}$ and 
$w_{\infty}$. Let us call their local neighbourhoods $U_z$ and
$U_w$, as local coordinates are  $z$ and $w$ respectively.  
At both asymptotic points the I-brane fermions will sweep out an
asymptotic state. The quantum partition function should therefore be
constructed from two quantum states. 

Before constructing these states for general $\la$, let us 
first consider the semi-classical limit $\la \to 0$. In this limit the
I-brane degrees of freedom are just conventional chiral fermions
on $\Sigma_{c=1}$.
The genus 1 part $\cF_1$ of the free energy is obtained as the
partition function of these semi-classical fermions. It can be
computed by assigning the Dirac vacuum $$| 0 \rangle_z = z^{1/2} \wedge
z^{3/2} \wedge z^{5/2} \wedge \ldots$$ to $U_z$ and likewise the conjugate 
state $$ |0 \rangle_w = w^{1/2} \wedge w^{3/2} \wedge w^{5/2} \wedge
\ldots$$ to $U_w$. To compare these states, we need an operator $S$
that relates $z$ to $1/z$. The semi-classical partition can then be
computed as a fermionic correlator 
$
_w \hspace{-.1mm} \langle 0 |S|0 \rangle_z,
$ 
with the result that 
\begin{align}\label{eq:F1c=1}
e^{\cF_1} =\, _w \hspace{-.2mm} \langle 0 |S|0 \rangle_z =  
\prod_{k\ge 0} \mu^{k+1/2}. 
\end{align}
Using $\zeta$-function regularization we find that this expression
yields the familiar answer $\cF_1 = -{1\over 12} \log \mu$.

In order to go beyond 1-loop, we should think in terms of
$\cD$-modules. Let us for a moment not represent their elements in
terms of differential operators yet. In both asymptotic regions we then
find the $\cD$-modules 
\begin{align*}
U_z: \quad &\cM = \cD / \cD P, \quad \textrm{with} \quad  P= \hat{z}
\hat{w} - \mu, \\ 
U_w: \quad &\underline{\cM} = \cD / \cD \underline{P}, \quad
\textrm{with} \quad 
\underline{P} =  \hat{w} \hat{z} - \mu + \la.
\end{align*}

Notice that the Weyl algebra $\cD = \langle \hat{z}, \hat{w} \rangle$,
with the relation $[\hat{z}, \hat{w}] = \la$, acts 
on monomials $z^k$ and $w^k$ in the module $\cM$ as   
\begin{eqnarray*}
 \hat{z} (z^k) = z^{k+1} && \hat{z} (w^k) =  \left( \la \d_w + \frac{\mu -
    \la}{w} \right) w^{k} \\ 
 \hat{w} (z^k) =  \left(- \la \d_z + \frac{\mu}{z}
\right)  z^{k}
&& \hat{w} (w^k) = w^{k+1}.  
\end{eqnarray*}

Here, we just used the relation $\cD P \equiv 0$ and wrote the elements
in the basis $\{ z^k, w^k \, | \, k \in \Z \}$ of $\cM$. A basis of a
representation of $\cM$ on which $\hat{z}$ and 
$\hat{w}$ just act by multiplication by $z$ resp. differentiation
with respect to $z$ is given by
\begin{align*}
&v^z_{k} (z) = z^k \cdot z^{-\mu/\la}, \\
&v^w_{k}(z) =  \int dw~e^{-zw/\la} ~ w^{k-1} \cdot w^{\mu/\la}.  
\end{align*}
Indeed, differentiation with respect to $z$ clearly gives the same
result as applying $\hat{w}$. Moreover, multiplying $v^w_{k}$ by
$z$ gives
\begin{align*}
z \cdot v^w_{k} (z) &= \la \int dw~ e^{-zw/\la}
\frac{\partial}{\partial w} \left( w^{k-1+ \mu/\la} \right) = (\mu
+\la( k -1)) v^w_{k-1}.
\end{align*}

Similarly, in the module $\underline{M}$ one can verify that 
\begin{eqnarray*}
 \hat{w} (w^k) = w^{k+1} && \hat{w} (z^k) =  \left( -\la \d_w  +
  \frac{\mu}{w} \right) w^{k} \\ 
 \hat{z} (w^m) =  \left(\la \d_z +
  \frac{\mu-\la}{z} \right)  z^{k}
&& \hat{z} (z^k) = z^{k+1}.  
\end{eqnarray*}
Hence in the
representation of $\underline{\cM}$ defined by
\begin{align*}
&\underline{v}^w_{k} (w) = w^{k-1} \cdot w^{\mu/\la}, \\
&\underline{v}^z_{k}(w) =  \int dz~e^{zw/\la} ~ z^k \cdot z^{-\mu/\la},
\end{align*}
$w$ and $\partial_w$ act in the usual way. 

Since we moved
over to representations of the $\cD$-module where the differential
operator acts as we are used to, the $S$ transformation, that connects
the $U_z$ and the $U_w$ patch and thereby exchanges $\hat{z}$ and
$\hat{w}$, must be a Fourier transformation. This is clear from the
expressions for the basis elements $w$ and $\tilde{w}$: $S$
interchanges $v^z_{k}(z)$ with
$\underline{v}^z_{k}(w)$, and $v^w_{k}(z)$ with $\underline{v}^w_{k}(w)$.
In total we thus find the $\cD$-module elements
\begin{align}\label{eq:c=1module}
U_z: \quad &v_k^{z}, ~{v}_k^{w} \\
U_w: \quad &\underline{v}_k^{w},~ \underline{v}_k^{z} \notag
\end{align}

Representing the $\cD$-module in terms of differential operators
of course gives the same result. A fundamental solution of $P \Psi(z)=0$ is
$\Psi(z) = z^{-\mu/\la}$, so that acting with $\cD =
\langle z, \partial_z \rangle$ on $\Psi(z)$ gives the elements $v^z_k$
in $\cM$. Likewise, we reconstruct the elements $\underline{v}^w_k$
from the fundamental solution of $\underline{P}
\underline{\Psi}(w)=0$. Since $\cD =\langle z, \partial_z \rangle$ and
$\underline{\cD} =\langle w, \partial_w \rangle$ are related by a
Fourier transform, an element $v_k$ of the $\cD$-module in one asymptotic
region is represented by its Fourier transform in the opposite
region. This reproduces all elements in 
(\ref{eq:c=1module}).



A $\la$-expansion of the $\cD$-module element  $\underline{v}^z_{k}$,
using for example the stationary
phase approximation, yields as zeroth order contribution
\begin{align*} 
e^{\mu/\la} \left( \frac{\mu}{w} \right)^{k-\mu/\la},
\end{align*}
while the subdominant contribution is given by 
\begin{align*} 
\sqrt{-\frac{2 \pi \la \mu}{w^2}}.
\end{align*}
So in total we find that 
\begin{align*}
\underline{v}^z_{k}(w) = \sqrt{-2 \pi \la}~(\mu/e)^{-\mu/\la}~ w^{\mu/\la}~
\mu^{k+1/2}~ w^{-k-1}~ \psi_{\textrm{qu}}\left(\frac{\mu}{w}\right).
\end{align*}
This summarizes the contributions that we found
before: the genus zero $w^{\mu/\la}$ and genus one
$\mu^{k+1/2} w^{-k-1}$ results, plus the higher  
order contributions that are collected in $\psi_{\textrm{qu}}$. 

The all-genus partition function $Z$ of this I-brane system can be
easily computed exactly. Schematically it equals the
correlation function     
\begin{align*}
Z_{c=1} =  \langle \cW_w | S_{\mu} |  \cW_z \rangle, 
\end{align*}
where the $S$-matrix implements the Fourier transform between the two
asymptotic patches. Similar to the arguments in (the appendices of)
\cite{Alexandrov:2002fh} and \cite{adkmv}\footnote{The
  argument presented in the appendix of \cite{adkmv} is not
  fully correct. The proper argument (as shown below) recovers a
  slightly different prefactor in front of the Gamma-function, related
  to the doubling in the appendix of \cite{Alexandrov:2002fh}.}  
we find that the
result reproduces the perturbative expansion of the free energy as in 
equation (\ref{eqn:Gaussianpartfunc}).
For completeness let us review the argument by comparing 
$\underline{v}^z_k(w)$ with $\underline{v}^w_k(w)$. 

Notice that $\underline{v}^z_{k}(w)$ almost equals the 
gamma-function $\Gamma(z) = \int_0^{\infty}
dt~e^{-t}~t^{z-1}$. 
Indeed, let us take the integration contour from $-
i\infty$ to $i\infty$ and choose the cut of the logarithm to run from
 $0$ to $\infty$. Then 
\begin{align*}
& \underline{v}^z_{k}(w) =  \left( \frac{\la}{w} \right) \int_{- i\infty}^{i \infty} dz'~e^{z'} ~
\left( \frac{\la z'}{w} \right)^{k-\frac{\mu}{\la}} \\
&= \left( \frac{i \la }{w}
\right)^{k+1-\frac{\mu}{\la}}  \left[ \int_{- \infty}^{0} dz'~e^{iz'}
  ~e^{(k- \frac{\mu}{\la}) \log z'}  + \int_{0}^{\infty} dz'~e^{iz'}
  ~e^{(k-\frac{\mu}{\la}) \log z'} \right] \\
& = \left( \frac{i \la }{w}
\right)^{k+1-\frac{\mu}{\la}}  \left[ \int_{i \infty}^{0} dz'~e^{iz'}
  ~e^{(k- \frac{\mu}{\la}) \log z'}  + \int_{0}^{i\infty} dz'~e^{iz'}
  ~e^{(k- \frac{\mu}{\la}) \log z'} \right],
\end{align*}
where we moved the contour along the positive imaginary axis. A change
of variables and using that
$\log(iz' - \epsilon) = \log z' - 3 i \pi/2$ and $\log(iz' + \epsilon) =
\log z' + i \pi/2$, for $\epsilon$ small and real, then yields   
\begin{align*}
&\underline{v}^z_k(w) =  \left( \frac{i\la }{w}
\right)^{k+ 1-\frac{\mu}{\la}} 
\left[ e^{\pi i (k +1- \frac{\mu}{\la})/2} -  e^{-3 \pi i (k +1 -
   \frac{\mu}{\la})/2} \right]
\Gamma \left( k + 1- \frac{\mu}{\la} \right). 
\end{align*}
which is the same as the theory of type II result in the appendix of
\cite{Alexandrov:2002fh}.\hyphenation{ig-no-ring} 
Ignoring the exponential factor (which will only play a role
non-perturbatively), we find that the free energy $\cF$ equals the sum 
\begin{align*}
 \cF \left( \la,\, \mu \right) &=  \sum_{k \ge 0}  \left(k+1 -{\mu \over
     \la}\right) \log  \la +  \log \Gamma \left( k + 1-
 \frac{\mu}{\la} \right). 
 \end{align*}
It obeys the recursion relation 
 \begin{align}
 \cF \left( \la,\, \mu  + {\la \over 2} \right) - \cF \left( \la, \, \mu
    - {\la \over 2} \right) =   \left( \frac{1}{2}  - {\mu
     \over \la}  \right) \log \la + \log
 \Gamma \left(  {1 \over 2} - {\mu \over \la}  \right). \notag 
 \end{align}
which is known to be fulfilled by the $c=1$ string (see for example
Appendix A in  \cite{nekrasov-okounkov}), up to a term $ -{1 \over 2}
\log (2 \pi \la)$ that can be taken care of  by normalizing the
functions $\underline{v}_k$. The same result is found when analyzing
the function ${v}_k$.


This concludes our discussion of the $c=1$ string. It is the first
$\cD$-module example 
where we see how to handle curves with two
punctures. The physical interpretation of the I-brane set-up
furthermore provides a check of our formalism. Moreover, this example agrees
with the claim that the $\cD$-module partition function should be
invariant under different parametrizations. Both the representation as
$c=1$ curve, $\Sigma_{c=1}: \,zw = \mu$, and that as a Gaussian matrix
model spectral 
curve, $\Sigma_{mm}: \,y^2  = x^2 + \mu$, yield the same partition function.

\section{Seiberg-Witten geometries}\label{sec:SW}

Many times $\mathcal{N}=2$ supersymmetric gauge theories proved to
provide an important theoretical framework for testing new ideas in
physics. It should be fair to say that the most important advances in
this context are the solution of Seiberg and Witten in terms of a
family of hyperelliptic curves, as well as the explicit solution of
Nekrasov and Okounkov in terms of two-dimensional partitions. 
In what follows we will provide a novel perspective on these
results, by wrapping an I-brane around a Seiberg-Witten curve. The
$B$-field on the I-brane quantizes the curve, and a fermionic state is
obtained from the corresponding $\cD$-module. As we will see,
this state sums over all possible fermion fluxes through the
Seiberg-Witten geometry, and may be interpreted as a sum over
geometries. First we briefly review 
the Seiberg-Witten and Nekrasov-Okounkov approaches.

The solution of the $U(N)$ Seiberg-Witten theory is encoded in its partition
function $Z(a_i,\lambda,\Lambda)$, which is a function of the scale
$\Lambda$, the coupling $\lambda$ and boundary conditions for the
Higgs field denoted by $a_i$ for $i=1,\ldots,N$ (with $\sum_i a_i=0$
for the $SU(N)$ theory). The partition function is related to the free
energy $\cF$ as 
\begin{align}
Z(a_i,\lambda,\Lambda) = e^{\cF} = e^{\sum_{g=0}^{\infty} \lambda^{2g-2} \cF_g(a_i,\Lambda)}.
\end{align}  
In the above expansion $\cF_0$ is the prepotential which contains in
particular an instanton expansion in powers of $\Lambda^{2N}$, while
higher $\cF_g$'s encode gravitational corrections. The $U(N)$ Seiberg-Witten
solution identifies the $a_i$'s and the derivatives of the prepotential
$\frac{1}{2\pi i} \frac{\partial \cF_0}{\partial a_i}$ as the $A_i$ and
$B_i$ periods of the meromorphic differential  
\begin{align}
\eta_{SW} = \frac{1}{2\pi i} v \frac{dt}{t}
\end{align}  
on the hyperelliptic curve 
\begin{align}
\Sigma_{SW} : \quad \Lambda^N (t + t^{-1}) = P_N(v) = \prod_{i=1}^N (v-u_i).   \label{SW-curve}
\end{align}  
Despite great conceptual advantages, extracting the instanton
expansion of the prepotential from this description is a non-trivial
task. However, an explicit formula for the partition function,
encoding not only the full prepotential but also entire expansion in
higher $\cF_g$ terms, was postulated by Nekrasov in \cite{Nek}.
Subsequently this formula was derived rigorously jointly by him and Okounkov in
\cite{nekrasov-okounkov} and independently by Nakajima and Yoshioka in \cite{NY-I,NY-II}. 
For $U(N)$ theory this partition function is given by a sum over $N$ partitions $\vec{ R}=(R_{(1)},\ldots,R_{(N)})$ 
\be \label{eqn:U(N)NOpartfunct}
Z(a_i,\lambda,\Lambda) =  Z^{\mathrm{pert}}(a_i,\lambda) \sum_{\vec{ R}} \Lambda^{2N |\vec{R}|} \mu^2_{\vec{R}}(a_i,\la), 
\ee
where
\begin{align}
\mu^2_{\vec{ R}}(a_i,\la) & =  \prod_{(i,m)\neq (j,n)} \frac{a_i - a_j + \la(R_{(i),m} - R_{(j),n} + n - m)}{a_i - a_j + \la(n-m)}, \label{muR}  \\
Z^{\mathrm{pert}}(a_i,\lambda) & =  \textrm{exp}\,\Big(\sum_{i,j} \gamma_{\lambda}(a_i-a_j,\Lambda)  \Big).   \label{Zpert}
\end{align}
The function $\gamma_{\lambda}(x,\Lambda)$ is related to the free energy of the topological string theory on the conifold, and its various representations and properties are discussed extensively in \cite{nekrasov-okounkov} in Appendix A.
The vevs $a_i$ are quantized in terms of $\lambda$, so that for $p_i\in\mathbb{Z}$,
\be
a_i = \lambda (p_i+\rho_i),\qquad \qquad \rho_i = \frac{2i-N+1}{2N}.
\ee 
The approach of \cite{Nek} is based on the localization technique in
presence of the so-called $\Omega$-background.
In general this background provides a two-parameter generalization of the
prepotential: the coupling $\lambda$ is replaced by two geometric
parameters $\epsilon_1$ and $\epsilon_2$. The prepotential, as given
above, is recovered for $\lambda=\epsilon_1=-\epsilon_2$. By the
duality web in $\S$\ref{ssec:web} supersymmetric gauge theories
are related to intersecting brane configurations. The
Nekrasov-Okounkov solution must therefore have an interpretation in
terms of a quantum Seiberg-Witten curve, where $\la$ plays the role of
the non-commutativity parameter.

\subsection{Dual partition functions and fermionic correlators}

For a relation to the I-brane partition
function (\ref{fermionpartfunc}), it is necessary to consider the dual 
of the partition function (\ref{eqn:U(N)NOpartfunct}). This is introduced in
\cite{nekrasov-okounkov} as the Legendre dual 
\be
Z^D(\xi,p,\lambda,\Lambda) = \sum_{\sum_i p_i = p} Z(\lambda(p_i+\rho_i),\lambda,\Lambda) \, e^{\frac{i}{\lambda}\sum_j p_j\xi_j}.  \label{Zdual}
\ee
An important observation of Nekrasov and Okounkov is that this dual
partition function can be elegantly written as a free fermion
correlator. This is a consequence of the correspondence between fermionic 
states and two-dimensional partitions described in Appendix~\ref{sec:infgrassmannian}. 
For $U(1)$ there is no difference between the partition function and its dual and both can be written as
\begin{align} \label{eq:fermcorr-u1}
Z_{U(1)}^{D} (p,\la,\Lambda) & = \langle p | e^{-\frac{1}{\la}
  \alpha_{1}} \Lambda^{2 L_0} 
 e^{\frac{1}{\la} \alpha_{-1}} |p \rangle,
\end{align}
where $|p \rangle$ is the fermionic vacuum whose Fermi level is raised
by $p = a/\la$ units and $L_0$ measures the energy of the state. A version of the
boson-fermi correspondence implies the following decomposition  
\begin{align}
 e^{\frac{1}{\la} \alpha_{-1}} |p \rangle = \sum_R
 \frac{\mu_R}{\la^{|R|}} |p;R \rangle              \label{u1-bos-fer}
\end{align}
in terms of partitions $R$, where $\mu_R$ is the Plancherel measure  
\begin{align}
\mu_R = \prod_{1\leq m<n <\infty} \frac{R_m - R_n + n - m}{n-m} = \prod_{\square \in R}\frac{1}{h(\square)}
\end{align}
which can be written equivalently as a product over hook lengths $h(\Box)$. 

For general $N$ the dual partition function (\ref{Zdual}) looks very similar 
\begin{align} \label{eq:fermcorr}
Z_{U(N)}^{D} (\xi_i;p,\la,\Lambda) & = \langle p | e^{-\frac{1}{\tilde{\la}}
  \alpha_{1}} e^{H_{\xi_i}} \Lambda^{2 L_0} 
 e^{\frac{1}{\tilde{\la}} \alpha_{-1}} |p \rangle, 
\end{align}
however, now this expression is obtained by blending $N$ free fermions
$\psi^{(i)}$ 
into a single fermion $\psi$, as explained in Appendix~\ref{sec:infgrassmannian}. 
In particular 
$H_{\xi_i} = \frac{1}{\la} \sum_r \xi_{(r+1/2) \hspace{-1mm} \mod N}
\psi_r  \psi^{\dag}_{-r}$, 
while the bosonic mode $\alpha_{-1}$ arises from the bosonization of
the single blended 
fermion $\psi$. In formula (\ref{u1-bos-fer}) the Plancherel measure
of a blended partition ${\bf R}$ 
can be decomposed into $N$ constituent partitions as
\be
\mu_{\bf R} = \sqrt{Z^{\mathrm{pert}}(a_i,\lambda)} \, \mu_{\vec{R}}(a_i,\lambda),   \label{Planch-Rvec}
\ee
with $\mu_{\vec{R}}$ and $Z^{\mathrm{pert}}$ given in (\ref{muR}) and (\ref{Zpert}). 
When read in terms of the $N$ twisted fermions $\psi^{(i)}$,  
the correlator (\ref{eq:fermcorr}) involves a sum over the individual fermion charges $p_i$. 


Our aim in this section is to derive the above fermionic expressions
for the dual partition function  
from the perspective of this paper.
In the next subsections we will see how canonically quantizing the
Seiberg-Witten curve in terms of a $\cD$-module elegantly reproduces
to the fermionic correlators (\ref{eq:fermcorr-u1}) and  
(\ref{eq:fermcorr}).

\subsection{Fermionic correlators as $\cD$-modules}

In this section we compute the I-brane partition function for $U(N)$
Seiberg-Witten geometries. We start with the simpler $U(1)$ and $U(2)$ examples
and then generalize this to $U(N)$. As a first principal step we
notice that the $U(N)$ Seiberg-Witten geometry 
\begin{align}
\Sigma_{SW} : \quad \Lambda^N (t + t^{-1}) = P_N(v) = \prod_{i=1}^N (v-u_i),  \label{uN-SWcurve}
\end{align}  
can be rewritten as 
\begin{align}
(P_N(v) - \Lambda^N t)(P_N(v) - \Lambda^N t^{-1})= \Lambda^{2N}.
\end{align}
This shows that the Seiberg-Witten surface may be seen as a transverse
intersection 
of a left and a right half-geometry defined by 
\begin{align}
\Sigma_{L}:~ \Lambda^N t = P_N(v) \quad
\mbox{resp.} \quad  \Sigma_{R}: ~\Lambda^N t^{-1} = P_N(v), 
\end{align}
which are connected
by a tube of size $\Lambda^{2N}$. The left geometry  parametrizes
the asymptotic region 
where both $t \to \infty$ and  $v \to \infty$, whereas the right
geometry describes the region where  $v
\to \infty$ while $t \to 0$. This is illustrated in figure~4.

\begin{figure}[h!]
\begin{center}   \label{fig4}
\includegraphics[width=5.7cm]{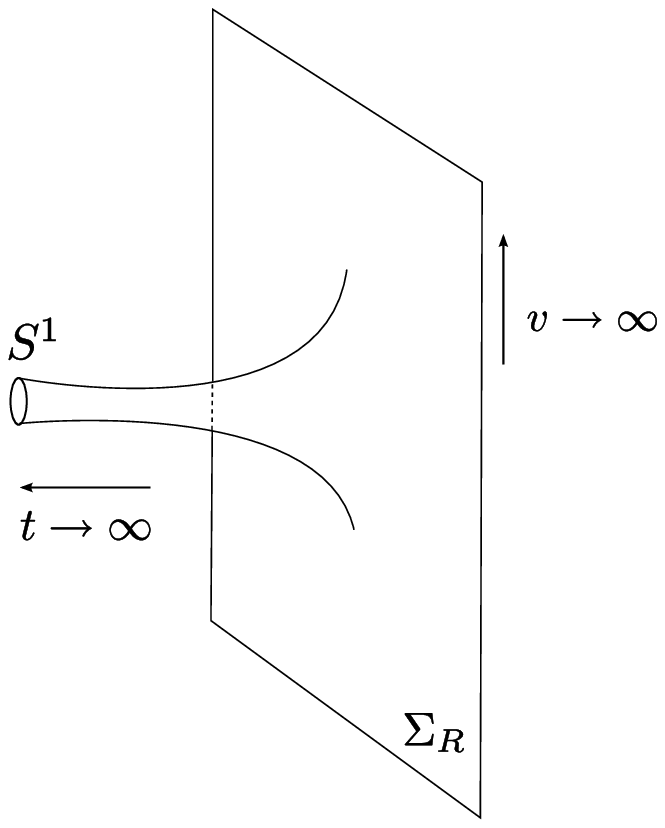}\\[5mm]
\parbox{12cm}{\small \bf Fig.\ 4: \it The right-half Seiberg-Witten
  geometry is distorted around the asymptotic point $(t \to 0, v
  \to \infty)$. A fermion field on the quantized curve can be
  described as an element of a $\cD$-module, and sweeps out a state
  $|\cW \rangle$ at the $S^1$-boundary where $t \to \infty$.}
\end{center} 
\end{figure}

Next we wish to associate a subspace in the Grassmannian to both half
Seiberg-Witten 
geometries. This will be swept out by a fermion field on the curve
that couples to the holomorphic part of the $B$-field
\begin{align}
B = \frac{1}{\la} ds \wedge dv
\end{align} 
Since this $B$-field quantizes the coordinate $v$ into the differential
operator~$\la \partial_s$, any subspace in this section is a
$\cD$-module for the differential algebra
\begin{align}  
D_{\C^*} = \langle t, \la \partial_s \rangle.
\end{align}

The free fermions on the Seiberg-Witten curves couple to the gauge field
$A = \frac{1}{\la} \eta_{SW}.$ This determines their flux through the
$A_i$ cycles of the Seiberg-Witten geometry as 
\begin{align}
p_i = \frac{1}{\la} \int_{A_i} \eta_{SW}.
\end{align}
The flux leaking through infinity is $p = \sum_{i=1}^N p_i$, which is
zero for $SU(N)$. A fermion field with fermion flux $p$ at infinity,
will sweep out a 
fermionic state in the $p$th Fock space. 
The parameters $\xi_i = \int_{B_i} \eta_{SW}$ are
dual to the fermion fluxes. Notice that in the perturbative regime $p_i$ can be
written as a $\la$-expansion 
\begin{align}   
 \la p_i =  u_i + \mathcal{O}(\la).
\end{align} 

Since both half Seiberg-Witten geometries are distorted near $v =
\infty$ (see figure~4), while a 
fermionic subspace can be read off in the neighbourhood where $v$ is
finite, both half-geometries parametrize a subspace of $\C((v))$: 
\begin{align}
\Sigma_{L}, ~\Sigma_{R} \subset \C((v)).
\end{align}
The trivial geometry corresponds to a disk with
origin at $v=\infty$, whereas its boundary encloses the
point $v=0$. The vacuum state is therefore given by 
\begin{align}\label{eqn:SWvacuum}
|0 \rangle = v^{-1/2} \wedge v^{-3/2} \wedge v^{-5/2} \wedge
\ldots.
\end{align}
Exponentials in $v^{-1}$ act trivially (as pure gauge
transformations in $ \Gamma_+$) on this  
state, whereas exponentials in $v$ transform the vacuum into a
non-trivial fermionic state.

Finally, the partition function is recovered by contracting the left
and the right fermionic state. 
Note that $s = - \log t$ is a local spatial coordinate on both half
Seiberg-Witten geometries, which tends to $-\infty$ on the left and to
$+ \infty$ on the right. This makes a huge difference with the $c=1$ geometry
discussed in \cite{adkmv,dhsv}, where the local coordinate is the
exponentiated coordinate, which on the left
is the inverse of that on the right. While in that example a non-trivial
$S$-matrix is required to identify the left and right half-geometries,
here we can just glue the fermionic states using the classic
Hamiltonian~$L_0$.  

Let us now find these quantum states! 

\subsub{$U(1)$ theory}

The $U(1)$ Seiberg-Witten curve is embedded in $\C^* \times \C$ as 
\bea
\Lambda (t+t^{-1}) = v - u, \qquad (t = e^s \in \C^*,~ v \in \C)
\eea
where $u \in \C$ is a normalizable mode. This geometry may be
factorized into a left and a right geometry        
\bea
\Sigma_L: ~ v = \Lambda t +  u \quad & \mbox{and} &  \quad \Sigma_R:
~ v = \Lambda t^{-1} + u,
\eea
that intersect transversely with degeneration parameter $\Lambda^2$.

The symplectic form $B = \frac{1}{\la} ds \wedge dv$ quantizes both
half geometries into $\cD_{\la}$-modules on a punctured disc $\C^*_t$,
parametrized by $t$. We claim that these are characterized by the
$U(1)$ $\la$-connections  
\bea
\nabla_L = - \la t \partial_t +  \Lambda t + \la p \quad & \mbox{and} &  \quad 
\nabla_R = \la t \partial_t  + \Lambda t^{-1} + \la p. 
\eea
These are just the canonical quantizations of the classical Seiberg-Witten
geometries, where additionally $u$ is quantized into $\la p$, with $p \in
\Z$. They yield the linear differential equations   
\begin{align}
P_L \psi^{\la}_{L}(t;p) &=  \left( - \la t \partial_t + \Lambda t +
  \la p  \right) \psi^{\la}_{L}(t;p)
 = 0,  \label{eqn:u(1)diffeqn} \\
P_R \psi^{\la}_{R}(t;p) &=  \left( \la t \partial_t + \Lambda t^{-1} +
  \la p \right)
\psi^{\la}_{R}(t^{-1};p)  = 0.
\end{align}
The $\cD_{\la}$-modules are of the canonical form
\begin{align}
\cM_{L/R} = \frac{\cD_{\la}}{ \cD_{\la}\cdot P_{L/R}},
\end{align}
and are generated by the solutions 
\bea
\psi_L^{\la}(t;p) = t^p e^{\frac{\Lambda}{\la} t} \quad & \mbox{and} & \quad
\psi_R^{\la}(t;p) = t^{-p} e^{\frac{\Lambda}{\la} t^{-1}}.
\eea
%

%
%


From the discussion in Appendix~\ref{sec:infgrassmannian} it follows that 
the factor $t^{-p}$ acts on the right Dirac vacuum by raising the Fermi level into $|p\rangle$, while the exponent of $ t^{-1}$ translates to the exponentiated $\alpha_{-1}$ operator. With an analogous statement for the left state, the modules $\cM_{L/R}$ translate into the Bogoliubov states
\bea
\langle \cW_L| = \langle p | e^{\frac{\Lambda}{\la}\alpha_1} \qquad &\mbox{and}& \qquad 
| \cW_R \rangle = e^{\frac{\Lambda}{\la}\alpha_{-1}}|p\rangle.    \label{u1-LR}
\eea

\begin{figure}[h!]
\begin{center}   \label{fig5}
\includegraphics[width=6cm]{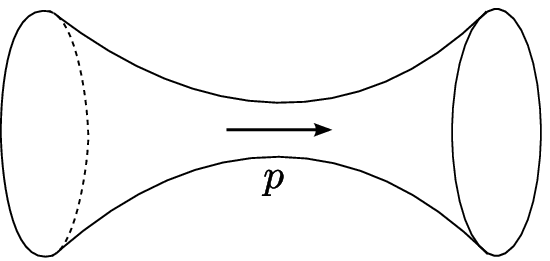}\\[5mm]
\parbox{12cm}{\small \bf Fig.\ 5: \it Contracting two
  Seiberg-Witten half-geometries yields the Nekrasov-Okounkov
  partition function corresponding to a fermion flux $p$ through the surface.}
\end{center} 
\end{figure}
 
The $U(1)$ Nekrasov-Okounkov partition function with fermion flux $p$ (see
figure~5) is found by contracting the above fermion states
\bea
Z^{\la}_{NO}(p;\Lambda) = \langle p | e^{\frac{\Lambda}{\la}\alpha_1}
e^{\frac{\Lambda}{\la}\alpha_{-1}}|p\rangle.  
\eea
The factors $\Lambda$ can be pulled out of the exponentials by using
the commutator $[L_0,\alpha_{\pm 1}] = \alpha_{\pm 1}$. Up to an extra
factor $\Lambda^{-p^2/2}$ we find that 
\bea
Z^{\la}_{NO}(p;\Lambda) \sim \langle p
|e^{\frac{\alpha_{1}}{\lambda}} 
\Lambda^{2 L_0} e^{\frac{\alpha_{-1}}{\lambda}} |p \rangle.  
\eea
This has a nice geometrical explanation, since the left and right half
geometries are connected by a tube of size
$\Lambda^2$ as in the factorized form of the complete $U(1)$
geometry. The factor $\Lambda^{2 L_0}$ is the Hamiltonian that
describes the propagation of the fermion field along the tube. There
is no need to 
generalize this standard-CFT factor, since both patches are described
by the same space-coordinate $s$.

We also note that, as consistent with \cite{adkmv}, the solution
$\psi^{\la}_{R}(t;u)$ to 
$P_R \psi = 0$ equals the one-point-function  
\bea
\langle p-1 | \psi(t) |\cW_R \rangle = \sum_n t^{-p-n} \langle
p;R_n|\cW_R \rangle = t^{-p} e^{\frac{\la}{\Lambda} t^{-1}} = 
\psi^{\la}_{R}(t;u), 
\eea
where $R_n$ represents a Young tableau consisting of just one row of $n$
boxes.

\subsub{$U(2)$ theory}

We apply now the above strategy for the $U(2)$ geometry. We split the corresponding curve into a left and
a right half geometry, and for brevity focus just on the right part defined by 
\bea
\Sigma_R: \quad \Lambda^2 t^{-1} = (v - u_2)(v - u_1).    \label{su2-right} 
\eea 
The $B$-field canonically quantizes this equation into the second order differential equation 
\bea \label{eqn:rank2swdiffeqn} 
P_{R} \psi(t) = \left\{ \la^2 (t \partial_t -  p_2 ) ( t \partial_t -
   p_1 ) -
  \Lambda^2 t^{-1} \right\}
\psi(s) = 0.
\eea
%

A change
of variables $z= 2 t^{-1/2}$
%
%
followed by the ansatz $\psi(z) = z^{-(p_1+p_2)} \phi(z)$ and the rescaling $z \mapsto
(\la/\Lambda)z$ transforms this differential equation into the
familiar Bessel equation  
\bea
\left( z^2 \partial_z^2 + z \partial_z -
  \nu^2 - z^2 \right) \phi(z) = 0, \quad \mbox{with} \quad \nu^2 = (p_1 - p_2)^2,
\eea
whose linearly independent solutions are given by modified Bessel
functions $I_{\nu}(z)$ and $K_{\nu}(z)$ of the first kind. The total solution in
the original $t$-coordinate is therefore a linear combination of
\begin{align}\label{eqn:solsu(2)sw}
\psi_R^{\la}(t;p_1,p_2) &= \left\{ \begin{array}{c}
t^{\frac{p}{2}} I_{\nu}\left( \frac{2 \Lambda}{\la \sqrt{t}} \right),\\  
t^{\frac{p}{2}} K_{\nu}\left( 
\frac{2 \Lambda}{\la \sqrt{t}}
\right), 
\end{array} \right.
\end{align}
where $p = p_1 + p_2$. These modified Bessel functions have different
asymptotics at infinity 
and relate to each other by going around the punctured disc $\C^*_t$.  

The second order differential operator $P_R$ defines the
$\cD_{\la}$-module 
\begin{align}
\cM_{R} = \frac{\cD_{\la}}{\cD_{\la} \cdot P_R},
\end{align}
which we claim represents fermions on the quantum $SU(2)$
Seiberg-Witten geometry.  
To check this statement, we have to find the fermionic state
corresponding to $\cM_R$. So we asymptotically expand of the
modified Bessel functions around $t=0$ in $\la$:
\begin{align*} 
I_{\nu}\left(\frac{2 \Lambda}{\la \sqrt{t}}\right) & \sim
t^{1/4}\exp \left(\frac{2
    \Lambda}{\la \sqrt{t}}\right)  \Big\{ 1 - 
\frac{(\mu -1)}{8}\, \frac{\la \sqrt{t}}{2
  \Lambda} + \frac{(\mu -1)(\mu-  
  9) }{ 2! \cdot 8^2} \, \frac{\la^2 t}{4
  \Lambda^2} + \ldots  \Big\}  \\
K_{\nu}\left(\frac{2 \Lambda}{\la \sqrt{t}}\right) &  \sim
t^{1/4}\exp \left(-\frac{2
    \Lambda}{\la \sqrt{t}}\right) \Big\{ 1 + 
\frac{(\mu -1)}{8}\, \frac{\la \sqrt{t}}{2
  \Lambda} + \frac{(\mu -1)(\mu-  
  9) }{ 2! \cdot 8^2} \, \frac{\la^2 t}{4
  \Lambda^2} + \ldots  \Big\},
\end{align*}
with $\mu = 4
\nu^2$.   

Recall that equation~(\ref{eqn:SWvacuum})  implies that any exponential
function in the local coordinate 
$v^{-1}=\sqrt{t}$ near the puncture acts trivially on the vacuum 
state. Equivalently, this is true for any asymptotic series in $\sqrt{t}$ that
assumes the value 1 at $\sqrt{t}=0$. In other words, we can
forget about the complete expansion in $\sqrt{t}$! Only the
WKB pieces
\begin{align}
t^{1/4} \exp \left(\pm {\frac{2 \Lambda}{\la \sqrt{t}}}\right)
\end{align}
are relevant in writing down the fermionic state. This is 
exactly opposite to the matrix model examples, where the
WKB-piece can be neglected and the perturbative series in $\la$
defines the fermionic state. 
 
The derivatives of the above solutions have one term
proportional to $\psi(s)$ (which we may forget about), and a term
proportional to the derivative of the Bessel functions. The latter may
be expanded as 
\begin{align*} 
\partial_s I_{\nu}(t) & \sim   t^{-1/4} \exp \left(\frac{2
    \Lambda}{\la \sqrt{t}}\right) \Big\{ 1 -
\frac{(\mu+3)}{8}\,  \frac{\la \sqrt{t}}{2
  \Lambda} + \frac{(\mu -1)(\mu+
  15)}{2! \cdot 8^2} \, \frac{\la^2 t}{4
  \Lambda^2}   + \ldots  \Big\}  \\
\partial_s K_{\nu}(t) & \sim t^{-1/4} \exp \left(\frac{2
    \Lambda}{\la \sqrt{t}}\right) \Big\{ 1 +
\frac{(\mu+3)}{8}\,  \frac{\la \sqrt{t}}{2
  \Lambda} + \frac{(\mu -1)(\mu+
  15)}{2! \cdot 8^2} \, \frac{\la^2 t}{4
  \Lambda^2}   + \ldots  \Big\}
\end{align*}
around $\sqrt{t}=0$. Again with the same reasoning only the WKB piece
is necessary to write down the quantum state. Taking into account the
extra factor $t^{\frac{p}{2}}$ in 
(\ref{eqn:solsu(2)sw}) the subspace $\cW^+_{R}$ is thus generated by the
$\cO(t)$-module  
\begin{align} 
t^{\frac{p}{2}} \left( \begin{array}{cc}  t^{\frac{1}{4}}  \exp \left(\frac{2
    \Lambda}{\la \sqrt{t}}\right) \\ 
      t^{-\frac{1}{4}}  \exp \left(\frac{2
    \Lambda}{\la \sqrt{t}}\right)\end{array} \right) \cO(t),
\end{align}
and blends (via the lexicographical ordening) into the fermionic  state 
\begin{align} 
| \cW^+_R \rangle = v^{-p} ~ e^{
      \frac{\Lambda}{\tilde{\la} }v} \left(  v^{\frac{1}{2}}  \wedge v^{-\frac{1}{2}}   \wedge   v^{-\frac{3}{2}} \wedge v^{-\frac{5}{2}} \wedge \ldots \right)
\end{align}
on the cover. Here we used a cover coordinate $v^{-1}$ obeying $v^{-2} =
t$, and rescaled the topological string coupling as $\tilde{\la} = \la/2$. $\cW^+_R$ is thus simply generated by a single function 
\begin{align} 
\psi^{\la}(v) = v^{-p} e^{\frac{\Lambda}{\tilde{\la}} v} 
\end{align}
Hence the fermions blend into the Bogoliubov state 
\begin{align} 
| \cW^+_{R} \rangle = e^{\frac{\Lambda}{\tilde{\la}} \alpha_{-1}} |p \rangle,   \label{su2-R}
\end{align}
when $p$ is an integer.

Note that the only modulus that appears in this expression is
$p$. This represents the diagonal $U(1)$, denoting the total fermion 
flux through the geometry. The moduli $p_1$ and $p_2$ measure the
fermion flux through an internal cycle and are not visible in the
result, because the final state sums over all internal
momenta. 
In general any $SU(2)$ Seiberg-Witten geometry with the same quantized
$p$ yields the same fermionic state.   

The fermionic (or dual) partition function is found by
contracting the left and the right states,  
similarly as in the $U(1)$ example above. The left state is just the
complex conjugate of the right one, so we find
\be
Z_{NO}^{D} (p;\la,\Lambda) = \langle p | e^{\frac{\Lambda}{\tilde{\la}} \alpha_{1}} 
 e^{\frac{\Lambda}{\tilde{\la}} \alpha_{-1}} |p \rangle  \sim \langle
 p | e^{\frac{1}{\tilde{\la}} \alpha_{1}} \Lambda^{2 L_0} 
 e^{\frac{1}{\tilde{\la}} \alpha_{-1}} |p \rangle.
\ee
The result is very similar to the $U(1)$ example, up to the shift $\la
\mapsto \la/2$. But notice that this fermionic state is written in
terms of a single blended fermion. Decomposing this fermion into two
twisted fermions makes it natural to insert an extra operator in the
middle of the correlator, that measures the momenta of the two
fermions through the $A$-cycles of the SW geometry. Weighting these
momenta with a potential $\xi_i$, for $i=1,2$, yields 
\begin{align} 
Z_{NO}^{D} (\xi_i,p;\la,\Lambda) & \sim \langle p | e^{\frac{1}{\tilde{\la}}
  \alpha_{1}} e^{H_{\xi_i}} \Lambda^{2 L_0} 
 e^{\frac{1}{\tilde{\la}} \alpha_{-1}} |p \rangle,
\end{align}
where $H_{\xi_i} = \frac{1}{\la} \sum_r \xi_{(r+1/2) \hspace{-1mm} \mod 2} \psi_r
\psi^{\dag}_{-r} = \frac{1}{\la} (p_1 \xi_1 + p_2 \xi_2)$. This is the
answer conjectured by Nekrasov and Okounkov in \cite{nekrasov-okounkov}.



\subsub{$U(N)$ theory}

It is not difficult to extend this discussion to the $U(N)$ 
theory (\ref{uN-SWcurve}), 
whose corresponding right half geometry we write as
\be
\Sigma_N :~ \Lambda^N t^{-1} = \prod_{i=1}^N
(v- u_i).   \label{uN-SWcurve-bis}
\ee
Canonically quantizing this geometry and changing the coordinates 
$z=  \left( \frac{\Lambda}{\lambda}
\right)^N t^{-1}$,
brings us to the degree $N$ differential equation
\be
P_N \psi(z) = \left( \prod_{i=1}^N (z\partial_z - p_i) - z \right) \psi(z) = 0.  \label{suN-eqn}
\ee

It turns out that a solution to the above equation is given by a
particular Meijer 
G-function, denoted $G^{m,n}_{p,q}(z)$. The Meijer G-function is a complicated special function which was introduced
in order to unify a number of standard special function
\cite{G-Meijer,G-Luke,G-Fields},
and is defined in terms of a complex integral
\begin{align}
G^{m,n}_{p,q}\left( \begin{array}{c}
a_1,\ldots,a_p \\
b_1,\ldots,b_q \end{array} | \, z \right) = \frac{1}{2\pi i} \int_L \frac{\prod_{j=1}^m \Gamma(b_j-t) \prod_{j=1}^n \Gamma(1-a_j+t)\, z^t }{\prod_{j=m+1}^q \Gamma(1-b_j+t) \prod_{j=n+1}^p \Gamma(a_j-t)} \, dt,
\end{align}
where $L$ is a contour which goes from $-i\infty$ to $+i\infty$ and
separates the poles of $\Gamma(b_j-t)$, for $j=1,\ldots,m $, from
those of $\Gamma(1-a_i+t)$, for $i=1,\ldots,n$.  

It can be shown that the Meijer G-function solves the differential equation
\begin{align}
\left( \prod_{i=1}^q (z\partial_z - b_i) + (-1)^{p-m-n+1}z
  \prod_{j=1}^p(z\partial_z - a_j + 1)\right) \,G(z) =
0. \label{Meijer-diffeq} 
\end{align}
So, indeed
the Seiberg-Witten differential equation (\ref{suN-eqn})  is a special
case of Meijer differential equation~(\ref{Meijer-diffeq})  with
$p=n=0$ and $q=N$. 
Therefore the differential equation (\ref{suN-eqn}) is solved by 
\begin{align}
\psi(z) =  G^{0,0}_{0,N}\left( \begin{array}{c}
\emptyset \\
p_1,p_2,\ldots,p_N \end{array} | \, z \right).
\end{align}

Similarly as before we claim that the $\cD$-module corresponding
to $U(N)$ Seiberg-Witten curve is generated by $P_N$. A subspace $\cW$
corresponding to this $\cD$-module is thus generated by a solution
$\psi(t)$ and its derivatives in $t \partial_t$. 

For $p<q$ the Meyer differential equation~(\ref{Meijer-diffeq}) has a
regular singularity at 
$z=0$ and an irregular one for $z=\infty$.
To extract the I-brane fermionic state, we are interested in
the behaviour around the irregular singularity, where $t
\to 0$.
It turns out that one of the independent solutions
of the Seiberg-Witten differential equation~(\ref{suN-eqn}) has the asymptotic
expansion \cite{G-Meijer,G-Luke,G-Fields} 
\begin{align}
\psi(v) \sim e^{\frac{\Lambda}{\lambda/N}v} \, v^{\frac{1-N}{2}}\, v^{p} \sum_{j=0}^{\infty} k_j v^{-j},
\end{align}
around this singularity,  which is conveniently written in
the cover coordinate $(-v)^N=t^{-1}=\left( \frac{\lambda}{\Lambda}   
\right)^N z$. The other solutions are found
by multiplying the coordinate $v$ by $N$-th roots of unity, and thus
behave distinctly at infinity. As before,
$p = \sum_{i=1}^N p_i$.  

To find the fermionic state corresponding to the $U(N)$ Seiberg-Witten curve,
we act with $\psi(v)$ on the Dirac vacuum. 
The positive power of $v$ in the exponent of $\psi(v)$ corresponds in
the operator 
language to $\alpha_{-1}$, whereas $v^p$ lifts the Fermi level.
The remaining series just contains negative powers of $v$ which translate to a
trivial action on the vacuum in the operator formalism. Therefore, the
above asymptotic solution and its derivatives (in $t \partial_t$)
blend into the state  
\be
|\mathcal{W}_R\rangle = e^{\frac{\Lambda}{\tilde{\lambda}}\alpha_{-1}} |p\rangle,     \label{suN-state}
\ee
with rescaled topological string coupling $\tilde{\lambda} = \lambda/N$.
Like for the $U(2)$ Seiberg-Witten geometry the dependence on the
individual moduli $p_{i}$ has dropped out. 

Similarly as in $U(1)$ and $U(2)$, in the present case we also find
the $U(N)$ Nekrasov-Okounkov dual partition function
\be
Z_{NO}^{D} (\xi_i;\la,\Lambda)  = \langle p | e^{\frac{1}{\tilde{\la}}
  \alpha_{1}} e^{H_{\xi_i}} \Lambda^{2 L_0} 
 e^{\frac{1}{\tilde{\la}} \alpha_{-1}} |p \rangle.\label{Z-NO}
\ee   
This
fermionic correlator is indeed the one postulated in \cite{nekrasov-okounkov}. For $N=1$ or $N=2$ the Meijer G-function specializes respectively to the
exponent and Bessel functions, which reproduces the results derived in
previous subsections.

Although the normalizable moduli $p_i$ disappear in the final I-brane partition
function, they reappear when the state is unblended in terms of $N$
single fermions
\begin{equation}\label{eq:microstates}
e^{\frac{1}{\tilde{\lambda}} \alpha_{-1}}|p\rangle = \sum_{ R} \frac{\mu_{ R}}{\tilde{\lambda}^{|{ R}|}} |p,{
R}\rangle = \sum_{ \sum p_i=p} \sum_{R_{(i)}} \sqrt{Z^{pert}(p)} \, \frac{\mu_{\vec{
R}}(p,\tilde{\lambda})}{\tilde{\lambda}^{|{ R}|}} \bigotimes_{l=1}^N |p_i,R_{(i)} \rangle,
\end{equation}
as may be seen from (\ref{u1-bos-fer}) and (\ref{Planch-Rvec}). The
charges $p_i$ have an interpretation as the fermion fluxes through the
$N$ tubes of  
the Seiberg-Witten geometry we started with. 

Actually, we find the same fermionic state when starting with any other
Seiberg-Witten geometry whose fermion flux at infinity is $p$. Hence
one microstate in the total sum~(\ref{eq:microstates}) can be
interpreted as a fermion flux through an infinite set of geometries. This
gives the state~(\ref{eq:microstates}) as well as the partition function
(\ref{Zdual}) the interpretation of a sum over geometries.

\subsection{Relation to  topological string theory}  \label{ssec:sum} 

Nekrasov and Okounkov also derive a partition function for
the 5-dimensional $U(N)$ Seiberg-Witten theory compactified on the
circle of circumference 
$\beta$ \cite{Nek,nekrasov-okounkov,NY-II} . It is given by a
$K$-theoretic generalization of the 4-dimensional formula in
equation (\ref{eqn:U(N)NOpartfunct}).

\begin{figure}[h!]
\begin{center}   \label{fig6}
\includegraphics[width=10cm]{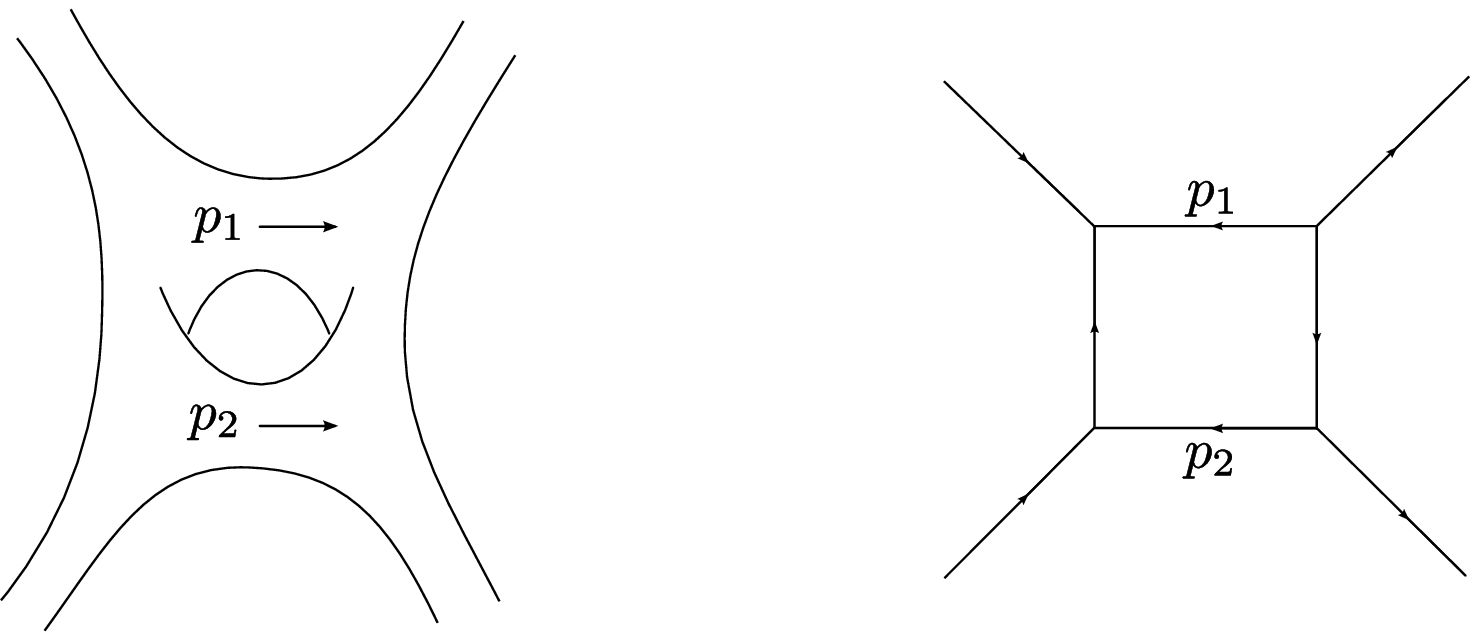}\\[5mm]
\parbox{12cm}{\small \bf Fig.\ 6: \it On the left we see the
  five-dimensional $U(2)$
  Seiberg-Witten surface with fermion fluxes through its
  $A$-cycles, and on the right a corresponding toric diagram. The
  fermion flux deforms the  K\"ahler lengths of the toric diagram as
  in equation~(\ref{kahlersize}).}
\end{center} 
\end{figure}

%
%

This 5-dimensional theory is closely related to the topological
string theory by geometric engineering \cite{KKV} on a toric
Calabi-Yau background \cite{Iq-KP-02,Iq-KP-03}. Namely, the
partition function of the topological string theory on an
$A_N$-singularity fibered over $\P^1$ (whose toric diagram consists of $N-1$
meshes as in figure~6) is equal to the partition function of the
5-dimensional gauge 
theory given above, when the 
K\"ahler sizes of the internal legs are 
\begin{align}\label{kahlersize}
Q_{F_i} = e^{\beta (a_{i+1} - a_{i} )}, \qquad  Q_B = \left( \frac{\beta \Lambda}{2} \right)^{2N}, 
\end{align}
where $F_i$ labels the vertical legs and $B$ the horizontal ones. 
In the so-called gauge theory limit, when $\beta\to 0$, the
topological string partition function reduces to the 4-dimensional
Seiberg-Witten partition function. The corresponding B-model 
mirror geometry is of the form 
\begin{align}
X_{SW}: \quad xy - H(t,v) = 0,
\end{align}
where $H(t,v)=0$ represents a Riemann surface
of genus $N-1$. In the gauge theory limit this surface
becomes the Seiberg-Witten curve $\Sigma_{SW}$, parametrized as in the
equation~(\ref{SW-curve}). 

In topological string theory it is natural as well to write down a dual
partition function \cite{adkmv}. In a local B-model this allows the
possibility of arbitrary fermion fluxes through the handles of the Riemann
surface. In this setting it has been argued before that turning on a
fermion flux is equivalent to deforming the geometry. More precisely,
fermion flux parametrized by $\mathcal{P} = p_i \cB_i$
changes the integral of the holomorphic 3-form over any linking
3-cycle $\cA_i$, 
%
%
and thereby shifts the complex structure moduli $S_i = \int_{\cA_i}
\Omega$ as
\begin{align}
S_i \mapsto S_i + \la p_i 
\end{align}
In the A-model fermion flux translates into wrapping D4 branes 
around 4-cycles, and thereby deforms the K\"ahler moduli. The I-brane
partition function 
thus equals the dual topological string partition function.  

\begin{figure}[h!]
\begin{center}   \label{fig7}
\includegraphics[width=10cm]{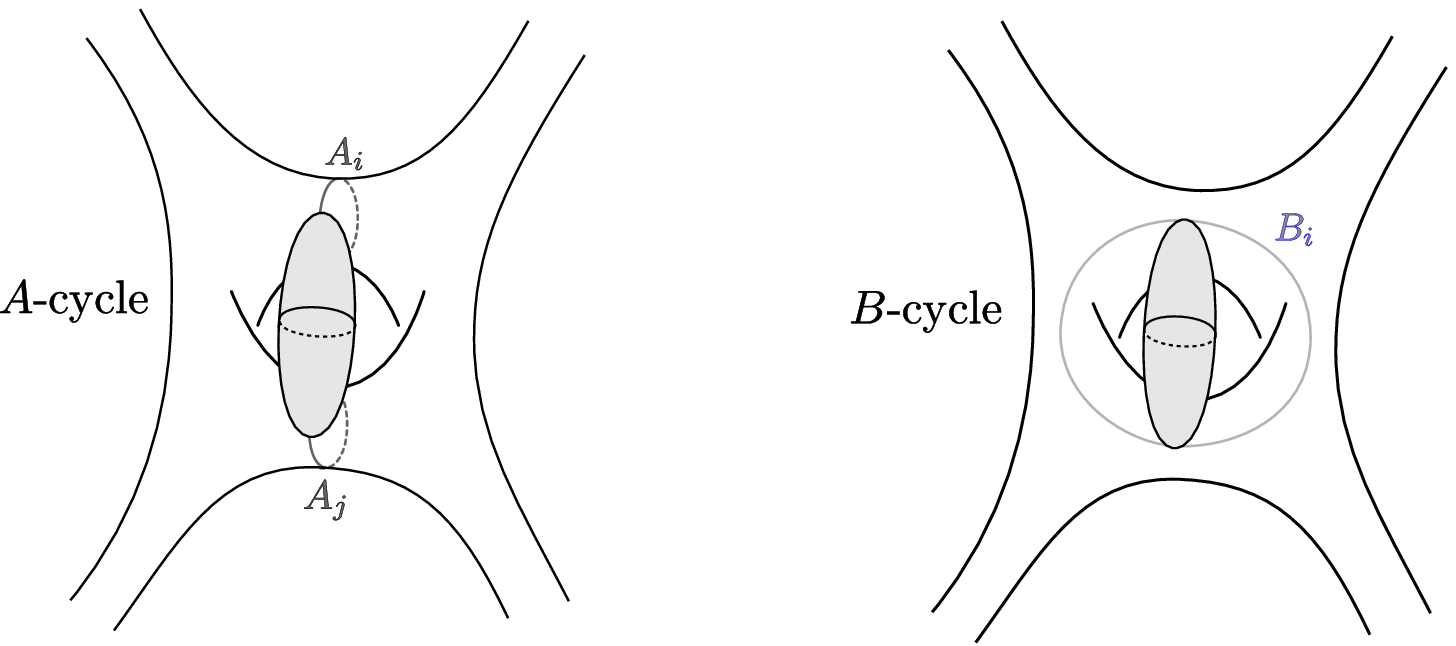}\\[5mm]
\parbox{12cm}{\small \bf Fig.\ 7: \it Three-cycles in the
  Seiberg-Witten $U(2)$-geometry.}
\end{center} 
\end{figure}
 
Because the Seiberg-Witten surface is embedded in $\C
\times \C^*$, $\cA$ and $\cB$-cycles in the toric threefold
will have topologies $S^1 \times S^2$ and $S^3$, respectively (they
are drawn in figure~7). In
particular, a basis of $\cA_i$-cycles can be chosen to reduce to the
surface as the combination $A_i - A_{i+1}$. Now notice that the
3-cycle $\cA_i$ with topology $S^1 \times S^2$ is mirror to the
vertical 2-cycle $F_{i}$ that connects the $i$-th and 
the $i+1$-th horizontal leg. So turning on a fermion flux $p_i$
through the $i$-th leg of the Seiberg-Witten geometry changes the
complex structure parameter $S_i$ by an amount proportional to $a_i -
a_{i+1}$. This explains the K\"ahler size $Q_{F_i}$ in (\ref{kahlersize}) 
in terms of fermionic fluxes through the Seiberg-Witten curve, and in
reverse why (\ref{eq:microstates}) may be interpreted as a sum over
Seiberg-Witten geometries, or equivalently toric diagrams.  
So we conclude that the fermionic interpretation in 4d of Nekrasov and
Okounkov is dual in 6d to the fermionic interpretation of the
topological string, and has a deeper interpretation in terms of $\cD$-modules. 

\subsub{Five-dimensional $U(1)$ theory}

Quantizing a five-dimensional Seiberg-Witten geometry yields
a difference (instead of differential) equation. Working out
$\cD$-modules for these geometries we leave for future work. Let us
treat one example in detail though. The five-dimensional right $U(1)$
Seiberg-Witten half-geometry 
\begin{align}\label{eqn:5dU(1)SW}
\Sigma^{5d}_R: \quad \beta \Lambda e^{-\beta \la} t^{-1} +
e^{-\beta v} 
-1 = 0 
\end{align}
may be drawn as a pair of pants. In the field theory limit $\beta \to 0$
it reduces to the familiar equation $\Lambda t^{-1} = v$ for the right-half
Seiberg-Witten geometry (with $u=0$). 

In the B-model the most general state
assigned to a local pair of pants geometry is given by a Bogoliubov state
\cite{adkmv} 
\be
|\cW \rangle  = \exp \Big[\sum_{i,j} \sum_{m,n=0}^{\infty} a^{ij}_{mn} \psi^i_{-m-1/2} \psi^{* j}_{-n-1/2}\Big]|0\rangle,  \label{V-state}
\ee 
where the index $i=1,2,3$ describes the fermion field on the three
asymptotic regions of the pair of pants, and the coefficients are
determined by a comparison with the A-model topological vertex. This
exponent can be expanded as a sum over states (see figure~8)  
\begin{align}
|p_1, R_1 \rangle \otimes |p_2, R_2 \rangle \otimes |p_3, R_3 \rangle,
\end{align}
where the fermion flux is conserved: $p_1 + p_2 + p_3 =0$. To describe
the 5d Seiberg-Witten $U(1)$ geometry we won't need this 
state in full generality. 

\begin{figure}[h!]
\begin{center}   \label{fig8}
\includegraphics[width=9.5cm]{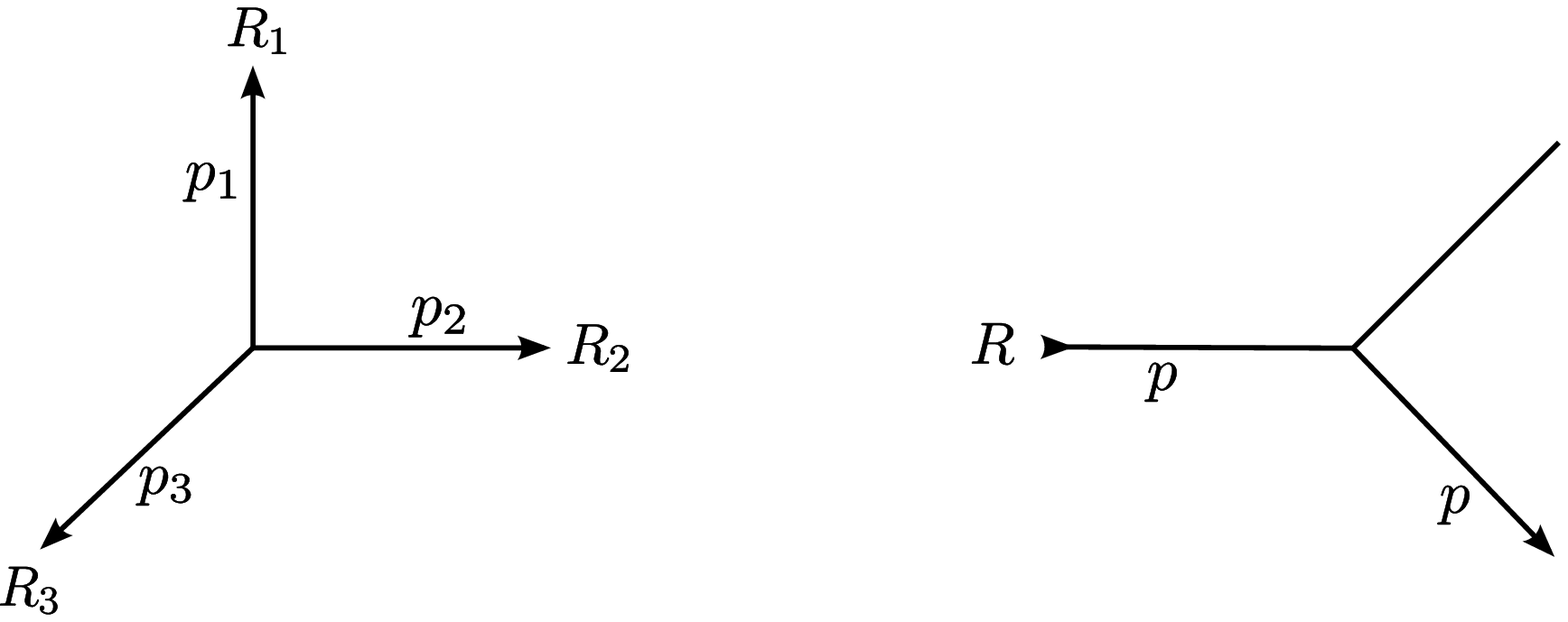}\\[5mm]
\parbox{12cm}{\small \bf Fig.\ 8: \it The B-model vertex (on the left)
  may be
  expanded as a sum over fermionic states $|p_1, R_1 \rangle \otimes
  |p_2, R_2 \rangle \otimes |p_3, R_3 \rangle$, with $p_1 + p_2 + p_3
  =0$, corresponding to a conserved fermion flux through the pair of
  pants. The five-dimensional right-half Seiberg-Witten geometry (on
  the right) with
  charge $p$ only has one partition $R \neq 0$. }
\end{center} 
\end{figure}

The B-field quantizes this geometry into the difference equation
\begin{align}\label{eq:difference-eqn}
P(t) \Psi(t)  = \left( \beta \Lambda e^{-\beta \la } t^{-1} + e^{\beta
 \la  t \partial_t} 
-1 \right) \Psi(t) =0.
\end{align}
%
Its fundamental solution is the quantum dilogarithm  
\begin{align}
\Psi(t) 
= 
\exp \sum_{n>0} \frac{ (\beta \Lambda)^{n} t^{-n} }{ n (1-e^{\beta \la n} ) }. 
\end{align}

As an intermezzo, notice that quantizing the equation
\begin{align}
\beta v = - \log \left( 1- \beta \Lambda e^{- \beta \la} t^{-1}
\right), 
\end{align}
which is just a rewriting of equation (\ref{eqn:5dU(1)SW}) for
$\Sigma^{5d}_R$, we 
find a differential equation which may be interpreted as the WKB
approximation of difference equation (\ref{eq:difference-eqn}). A
fundamental solution of the 
differential equation is given by the genus~0 disc amplitude 
\begin{align}
\Psi_0(u) 
= \exp \sum_{n>0} \frac{ (\beta \Lambda)^n t^{-n} }{ \la n^2 e^{\beta
    \la n} }. 
\end{align}

Acting with the five-dimensional dilogarithm on the Dirac vacuum state
yields the fermionic state 
\begin{align}
| \cW \rangle^{5d}_{U(1)} = \exp  \sum_{n>0}
\frac{ (\beta \Lambda)^{n} \alpha_{-n}}{n(1-e^{\beta \la n})}|0 \rangle.
\end{align}
This describes a subset of $| \cW
\rangle$ where only the quantum number $R_1$ is non-trivial. 
Summing over all external states of the form 
\begin{align}
|-p,R \rangle \otimes |p,\bullet \rangle \otimes |0, \bullet \rangle,
\end{align}
incorporates a fermion flux $p$ through the pair of pants. In the field
theory limit $\beta \to 0$ the resulting state reduces to the familiar
four-dimensional state $$\exp( \alpha_{-1}/\la ) |p \rangle
\otimes|p,\bullet \rangle \otimes |0, \bullet \rangle.$$

The partition function is found as the contraction of the left and
right 5d half-geometries. (Or equivalently in the topological B-model by
inserting a propagator \cite{adkmv}.) This yields the fermionic correlator 
\begin{align}
\langle 0| \tilde{\Gamma}_+ \tilde{\Gamma}_-|0 \rangle  
= \langle 0| \Gamma_+ (\beta \Lambda)^{2L_0} \Gamma_-| 0 \rangle, 
\end{align}
with 
\begin{align}
\tilde{\Gamma}_{\pm}=\exp  \sum_{\pm n>0}
\frac{ (\beta \Lambda)^{|n|} \alpha_{n}}{|n|(1-e^{\beta \la n})}
\quad \mbox{and} \quad
 \Gamma_{\pm}=\exp\sum_{\pm n>0}
\frac{ \alpha_{n}}{|n|(1-e^{\beta \la n})}.
\end{align}
Indeed, the result equals the five-dimensional $U(1)$ partition function 
\begin{align}
Z^{U(1)}_{5d}(\lambda, \Lambda, \beta) = \textrm{exp}\sum_{n=1}^{\infty}
\frac{(\beta \Lambda)^{2n}}{4n\,\textrm{sinh}^2(\beta \la n/2)},
\end{align}
that was found by Nekrasov and Okounkov in \cite{nekrasov-okounkov}.

\section{Discussion}\label{sec:discussion}

In this paper we argued that the fundamental objects underlying various systems in theoretical physics are chiral fermions living on quantum  curves. In our formulation the quantum curve is defined, similarly to an affine classical curve, in terms of an equation of the form $P(z,w)=0$. Its crucial feature, however, is the non-commutative character of the coordinates $z,w$. These quantum (or non-commutative) curves generalize the classical curves that come up in the standard formulation of a given topic. Examples of such classical curves can be found in the theory of random matrices, $c=1$ string theory, Seiberg-Witten theory, and more generally in topological string theory. Semi-classically their (genus one) free energy is computed as a fermionic determinant on the classical curve. In our approach chiral fermions on the quantum curve generate the all-genus expansion of the free energy with respect to the non-commutativity parameter~$\la$.  

Fermions on a non-commutative curve can be realized physically within string theory as massless states of open strings on an intersecting brane configuration in the presence of a $B$-field. This idea was already put forward in \cite{dhsv}. In this paper we have exploited this I-brane system in a few important examples. 

First of all we showed, reinterpreting the results in \cite{eynard-isomonodromic}, that I-branes and $\cD$-modules provide an insightful formulation of matrix models. This quite general statement is also appealing when certain matrix model limits are considered, such as a double scaling limits. In this case one recovers an I-brane formulation of minimal string theory and topological gravity. Secondly, we discussed how to reformulate $c=1$ string theory in the framework of $\cD$-modules. 

Finally, we discussed supersymmetric gauge theories.
Using $\cD$-module formalism we derived fermionic expressions for the partition function of the $\mathcal{N}=2$ gauge theory, reproducing the dual all-genus partition function introduced in \cite{nekrasov-okounkov}.
We considered mainly 4-dimensional Seiberg-Witten geometries with unitary gauge groups, and explained only the simplest $U(1)$ example in the 5-dimensional theory. It would be insightful to extend these results to other gauge groups and include matter content. It is clear that this should be possible, as these aspects of the 5-dimensional Seiberg-Witten theory are captured by topological string theory on toric manifolds. 
The latter system can be solved in fermionic B-model formulation of the topological vertex \cite{adkmv} which is equivalent to the I-brane fermions \cite{dhsv}. Nonetheless, finding the quantum I-brane curve representing such configurations appears to be a nontrivial task.  \\  


In all these examples we were able write down a $\cD$-module that, through the prescription in section~\ref{sec:formalism}, yields the all-genus partition function. Especially the matrix model examples made it clear that this $\cD$-module can be quite non-trivial in general. Only for the simplest curves, such as those appearing in double scaled matrix models, the $\cD$-module can be found by canonically quantizing the classical spectral curve.  

In the process of unraveling the $\cD$-module structure in both sets of examples, we noticed some crucial differences. While the WKB piece of the $\cD$-module generator can be ignored in finding the all-genus matrix model partition function, we discovered that it plays an eminent role for the Seiberg-Witten geometries. Another distinction is the difference in (non-)normalizable modes. While the potential $W$ parametrizes non-normalizable modes that appear in the $\cD$-module as parameters, in contrast, the normalizable modes in the Seiberg-Witten geometries are eaten by the $\cD$-module, and only visible as a sum over internal fermion fluxes in the geometry. On the other hand, varying the $\cD$-module with respect to the non-normalizable modes yields differential equations which relate to isomonodromy and the Stokes phenomenon.

Even with this rather broad set of examples, a few major questions remain. First of all, we cannot give a recipe in general how to find the quantum curve underlying a certain problem. Secondly, it is not obvious that our prescription is independent of the chosen parametrization of the classical curve. As we noted in the example of the classical curve $zw = 1$, different parametrizations can lead to different quantum curves that nonetheless yield the same partition function. This should hold in general cases as well, as topological string theory associates a unique all-genus partition function to a given curve. Thirdly, we haven't exploited some of the advantages of using $\cD$-modules instead of differential equations. One of the main advantages is some independence on the way the differential equation is written down. It would be very interesting to try to match this freedom with the choice of parametrization for the classical curve. Finally, we have only discussed examples with one or two local patches. It would be highly insightful to study more general examples.\\

While in this paper our focus has been to associate a $\la$-perturbative quantum state to a spectral curve, we noticed that $\cD$-modules in fact contain non-perturbative information. These bits get lost when we turn the $\cD$-module in a fermionic state by making an asymptotic expansion of the $\cD$-module generators in $\la$. This is in line with the discussion in \cite{Maldacena:2004sn}, where it is argued that non-perturbative effects drastically modify the non-trivial target space curve into a complex plane. Non-perturbative effects in matrix models, as well as in the topological string theory, were also recently discussed in \cite{Marino-nonpert1,Marino-nonpert2}. Especially interesting in this respect is \cite{Eynard:2008he}, where a non-perturbative partition function is proposed that is very similar to the I-brane partition function (\ref{fermionpartfunc}).

In the step where we turn a $\cD$-module in a quantum state, a choice of boundary conditions has to be made. This implies that final states
are troubled by the Stokes effect: solutions that decay faster can be added at no cost and the state changes when one crosses certain lines
in the moduli space. This suggests that the $\cD$-modules we studied in this paper may help in the understanding of wall-crossing phenomena in the corresponding $\mathcal{N}=2$ theories \cite{denef-moore, gaiotto-moore-neitzke}. \\ 

More mathematically, our formalism is deeply connected with quantum integrable systems and the geometric Langlands program \cite{frenkellectures, langlands, witten-ramification, beilinsondrinfeld, talkarinkin, chervov-talalaev, branes-quant}, while approaching these topics from a string theoretic perspective. Especially interesting in this respect is our quantitative approach, that allows us to associate quantum invariants to spectral curves. In the future we hope to make this link even more concrete.

Specifically, it would be enlightening to have a better description of the non-commutative fermionic CFT on a given quantum curve. It is interesting to find out whether this relates to the WZW models based on opers in the geometric Langlands program: as so-called Hecke eigensheafs these generate examples of the Langlands correspondence. And, to discover the relation with the interacting bosonic CFT's that give another perspective on these intersecting brane configurations \cite{2dKS,eynardorantin,remodelingBmodel} as well as \cite{eynard-latest}. In particular, both models determine a set of recursion relations. It would be helpful to compare them. 


A clear physical realization of quantum curves and the associated well-defined mathematical formulation in terms of $\cD$-modules are great advantages of our approach. In consequence it can be applied to numerous situations mentioned above and yields definite quantitative results. Nonetheless, the idea of quantum curves is not new and earlier attempts of their formulation appeared before in physics and mathematics. It is worthwhile to recall how those attempts relate to our formalism.


The notion of quantum or non-commutative geometry has also been introduced by A. Connes \cite{connes}. His approach relies on replacing the algebra of functions on a manifold by a non-commutative $C^*$-algebra. In this context a program of developing a theory of non-commutative Riemann surfaces, from the point of view of geometric quantization \cite{Brezin}, was advanced in \cite{KL}. Independently of this program, also some particular examples of low genus non-commutative Riemann surfaces have been analyzed in literature. In genus zero they include the so-called Podle\'s sphere \cite{Podles} and more generally fuzzy spheres \cite{fuzzy}, which also found vast application in string theory. In genus one, one can consider a non-commutative torus which arises naturally in a certain realization of M-theory known as Matrix theory \cite{qtorus1,qtorus2}. Non-trivial $B$-field is an essential ingredient in a realization of these systems. It would be interesting to see if they could be related to I-brane configurations.

\vspace{12mm}

\centerline{\bf Acknowledgments} 

\bigskip

We would like to thank the Gauge Theory and Langlands Duality workshop
at KITP at the University of California at Santa Barbara for excellent
lectures, inspiring surroundings and enlightening discussions. In
addition P.S. highly appreciates the hospitality of the String Theory
group at the University of Amsterdam, the High Energy Theory group at
the University of California San Diego, and the $6^{th}$ Simons
Workshop in Mathematics and Physics at the Stony Brook University,
were parts of this work were done.  We especially thank D.~Arinkin,
D.~Ben-Zvi, B.~Eynard, E.~Frenkel, A.~Klemm, M.~Kontsevich, T.~Pantev
and C.~Vafa for discussions. The research 
of R.D. and L.H. is supported by a NWO Spinoza grant and the FOM
program {\it String Theory and Quantum Gravity}.  The research of
P.S. is supported by the Humboldt Fellowship.  This research is also
supported in part by DARPA and AFOSR through the grant
FA9550-07-1-0543 and by the National Science Foundation under Grant
No. PHY05-51164.

 \appendix

\section{Infinite dimensional Grassmannian}\label{sec:infgrassmannian}

In this section we introduce an infinite dimensional Grassmannian and
its description in terms of the second quantized fermion field
(we learned this material e.g. from \cite{segalwilson, satograssmannian,
  robbertcargese, mulaseKP, KacvdLeur}).

\subsub{Grassmannian and second quantized fermions}

The space $\cH = \C((z^{-1}))$ of all formal Laurent series in $z^{-1}$ can be
given an interpretation of a Hilbert space. Basis vectors $z^n$, for
$n\in \mathbb{Z}$, correspond to one particle states of energy $n$
associated to the Hamiltonian $z\partial_z$.  
This Hilbert space has a decomposition 
\begin{align}
\cH = \cH_+ \oplus \cH_-,     
\end{align}
such that the first factor $\cH_+=\C[z]$ is a subspace generated by
$z^0$, $z^1$, $z^2$, $\ldots$, while $\cH_-$ is generated by negative
powers $z^{-1},z^{-2},\ldots$. Consider now a subspace $\cW$ of $\cH$
with a basis 
$\{ w_k(z) \}_{k\in \mathbb{N}}$. 
We say it is comparable to
$\cH_+$, if in the projection onto positive and negative modes 
\be
w_k = \sum_{j\geq 0} (w_+)_{ij}z^j + \sum_{j>0}(w_-)_{ij}z^{-j}
\ee
the matrix $w_+$ is invertible. The Grassmannian $Gr_0$ is the set of all subspaces $\cW \subset \C((z))$ which are comparable to $\cH_+$.

In what follows we take much advantage of the
correspondence between $Gr_0$ and the charge zero sector of the second
quantized fermion Fock space $\cF_0$. In this correspondence the
subspace $\cH_+$ is quantized as the Dirac vacuum
\begin{align}
|0 \rangle = z^0 \wedge z^1 \wedge z^2 \wedge \ldots,    \label{wedge-vac}
\end{align}
with all positive energy states filled. The fermionic state associated to the
subspace $\cW$ with basis $w_0(z)$, $w_1(z)$, $w_2(z)$, $\ldots$ is
represented by the semi-infinite wedge\footnote{Actually, we have to tensor with $z^{\hf}$ to make the state fermionic.}
\begin{align}
|\cW \rangle = w_0 \wedge w_1 \wedge w_2 \wedge \ldots
\end{align}
which is an element of the fiber of a determinant line bundle over the
element $\cW \in Gr$ (and therefore determined up a complex scalar $c$).

To make contact with the usual formulation of the second quantized fermion Fock space, we can identify the differentiation and wedging operators with the fermionic modes 
\begin{align}
\psi_{n+\hf} = \frac{\partial}{\partial z^{-n}} \qquad  \psi_{n+\hf}^* = z^{n} \wedge. 
\end{align}
These half-integer modes are annihilation and creation operators which arise from a decomposition of the fermion field $\psi(z)$ and its conjugate $\psi^*(z)$
\be
\psi(z) = \sum_{r \in\Z + \hf} \psi_{r} z^{-r-\hf} \qquad \psi^*(z) = \sum_{r\in\Z+\hf} \psi^*_{r} z^{-r-\hf},   \label{fermion-NS}
\ee
and they obey the anti-commutation relations 
$
\{\psi_{r},\psi^*_{-s}\} = \delta_{r,s}.
$

For subspaces $\cW \in Gr_0$ the determinant of the projection onto $\cH_+$ is well defined and can be expressed as
\begin{align}
\det w_+ = \langle 0 | \cW \rangle. 
\end{align} 

More generally, one can consider the Fock space $\cF$ which splits into subspaces of charge $p$  
\be
\cF = \bigoplus_{p\in\Z}\, \cF_p.
\ee
Each subspace $\cF_p$ is built by acting with creation and annihilation operators on a vacuum 
\be
|p \rangle = z^p\wedge z^{p+1}\wedge z^{p+2}\wedge\ldots,
\ee
with the property
\bea
\psi_{r} |p\rangle & = & 0 \qquad \textrm{for}\ r > p, \nonumber \\
\psi^*_{r} |p\rangle & = & 0 \qquad \textrm{for} \ r > -p.
\eea
The Fermi level of the vacuum $|p \rangle$ is shifted by $p$ units
with respect to the Dirac vacuum $|0 \rangle$. 
This fermion charge is measured by the $U(1)$ current
\be
J(z) = :\psi(z)\psi^*(z): = \sum_n \alpha_n z^{-n-1},
\ee
whose components $\alpha_n = \sum_k : \psi_r \psi^*_{n-r}$ satisfy the bosonic commutation relations 
\be
[\alpha_m, \alpha_{-n}] = m \delta_{m,n}.
\ee
With each subspace  $\cW \subset \C((z))$ comparable to the one generated by $(z^k)_{k\geq p}$ one can associate a state $|\cW\rangle \in \cF$ of charge $p$. This charge is equal to the index of the projection operator $pr_+ : \cW \to \cH_+$.

\begin{figure}[h!]
\begin{center}   \label{fig3}
\includegraphics[width=6.5cm]{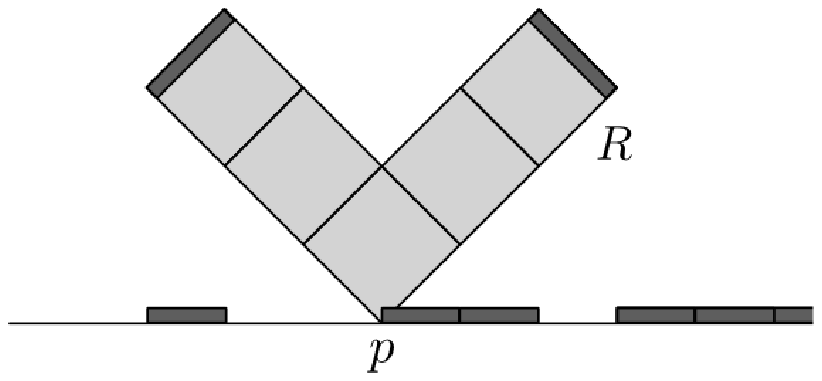}\\[5mm]
\parbox{12cm}{\small \bf Fig.\ 9: \it Elements of the Fock space
  $\mathcal{F}$  are in a bijective correspondence
  with Maya diagrams. The bottom line represent a Maya diagram
  corresponding to a fermionic state with charge $p$.
  As illustrated it is characterized by a two-dimensional partitions $R$ located
  at position $p$. We therefore denote the state as $|p,R\rangle \in \mathcal{F}$.} 
\end{center} 
\end{figure}

A state in the Fock space $\cF$ has also a simple representation in terms of the so-called Maya diagram (see figure~9). Black boxes in such a diagram represent excitations, whereas white boxes are gaps in the energy spectrum of the fermion. The charge of a state is given by the number of excitations minus the number of gaps. Fermionic states or Maya diagrams of a fixed charge $p$ can also be associated to two-dimensional partitions. In particular in $p=0$ sector the state
\be
|R\rangle = \prod_{i=1}^{d} \psi^*_{-a_i-\hf} \psi_{-b_i-\hf}|0\rangle
\ee
corresponds to the partition $R=(R_1,\ldots,R_l)$ such that
\be
a_i = R_i-i,\qquad b_i = R_i^t-i.
\ee
In what follows a state corresponding to a partition $R$ of charge $p$ is denoted as $|p,R\rangle$.

\subsub{Flow on the Grassmannian}

There is an action on the Grassmannian defined by multiplying a
basis vector $w_k(z)$ of $\cW$ by a power series $f(z)= \sum f_n
z^n$ that vanishes at $z=0$.
\bea
f(z) |\cW \rangle = \sum_k w_0 \wedge \ldots \w_{k-1} \wedge
f \cdot w_k
\wedge w_{k+1} \ldots .
\eea
When we write $w_k(z)$ in terms of the basis $(z^l)_{l \in \Z}$ this
action is encoded by the multiplication by an infinite matrix in
$gl_{\infty}$, whose $(i,j)^{th}$ entry is given by $f_{i-j}$. 
On the fermionic state $|\cW \rangle$ a multiplication by $z^n$ translates
into a commutator with the bosonic mode $\alpha_n$, since $\alpha_n$
increases the fermionic mode number by 
\be
[\alpha_n, \psi_r] =
\psi_{r+n}.
\ee 
Multiplication by a power series $f(z)$ therefore
translates to the operator 
\bea
f = \sum_n  f_n  [ \alpha_n, \bullet ]~ \in gl_{\infty}
\eea
on the Fock space.  

Exponentiating the action of $gl_{\infty}$ yields the group
$Gl_{\infty}$. An element $g(z) = \exp(f(z))$ of this group acts on
$|\cW 
\rangle$ by multiplying all its basis vectors   
\bea
g(z) |\cW \rangle = g \cdot w_0 \wedge \ldots \wedge g \cdot  w_k \wedge \ldots .
\eea
From the fermionic point of view this action is given by conjugating
each basis vector $w_k$ with the element   
\begin{align}
 g = \exp\left(\sum f_n \alpha_n \right)= \exp \left( \oint dz ~f(z) J(z) \right)
 \in Gl_{\infty}.
\end{align}

We call $\Gamma$ the group of exponentials $g(z) : S^1 \to \C^*$. 
An important subgroup of $\Gamma$ is the group $\Gamma_+$ of functions
$g_0: S^1 \to \C^*$ that extend over the disk $D_0 = \{z : |z| \le 1 \}$: 
\begin{align}
\Gamma_+ = \{ g_0: D_0 \to \C^* : g_0(0) =1 \}.   \label{Gplus}
\end{align}
Another subgroup is the group $\Gamma_-$ of
functions $g_{\infty}: S^1 \to \C^*$ that extend over the disk
$D_{\infty} = 
\{z \in \C \cup \{ \infty \}: |z| \le 1 \}$: 
\begin{align}
\Gamma_- = \{ g_{\infty}: D_{\infty} \to \C^* : g_{\infty}(\infty) =1 \}.
\end{align}
Any $g \in \Gamma$ can
be written as an exponential $\exp(f)$. When $g \in \Gamma_+$ the
function $f$ vanishes at $z=0$, and when $g \in \Gamma_-$ it
vanishes at $z=\infty$. 
 
$\Gamma_+$ and $\Gamma_-$ have different properties when acting on
Grassmannian.  
The action of $\Gamma_-$ is free, since any $\cW
\in Gr$ has only a finite number of excitations. On the contrary,
$\Gamma_+$ acts trivially on a vacuum state $|p \rangle$. 
Although the action of the groups $\Gamma_+$ and $\Gamma_-$ on a
subspace $\cW$  
is commutative, as it is just given by multiplication, as operators
on the fermionic state $|\cW \rangle$ it matters which element is
applied first. This introduces normal ordering ambiguities. 

An element  
\begin{align}
g(t,z) = \exp \left( \sum_{k \ge 1} t_k z^k \right) = \exp \left(
  f(t,z) \right) \in \Gamma_+,    \label{gexp}
\end{align}
defines a linear flow over the Grassmannian $Gr$. On the Fock space it
acts as an evolution operator 
\begin{align}
U(t) = \exp \left(\oint\frac{dz}{2\pi
    i} f(t,z)J(z)\right).
\end{align}
The determinant
det$(\cW)_+$ is not equivariant with respect to the action of $\Gamma_+$.
The difference is measured by the so-called tau-function 
\begin{align}\label{eqn:tauferm}
\tau_{\cW}(g)  = \frac{\det\,(g^{-1}w)_+} {g^{-1} \det\, w_+} =
\frac{\langle 0 | U(t) | \cW \rangle } {g^{-1} \langle
  0 | \cW \rangle},
\end{align}
which yields a holomorphic function $\tau: \Gamma_+ \to \C$.
This can be regarded as a wave function of $|\cW\rangle$.


%
%

%
%
%
%

\subsub{Blending}

So far we considered the Hilbert space $\cH\equiv \cH^{(1)}$ of
functions with values in $\C$. More generally, one can consider a
Hilbert space $\cH^{(n)}$ of functions with values in $\C^n$. Let
$(\epsilon_i)_{i=1,\ldots,n}$ denote a basis of $\C^n$. For each $n$
there is an isomorphism between $\cH^{(n)}$ and $\cH$ given by the
lexicographical identification of the basis
\be
\epsilon_i z^k \mapsto z^{nk+i-1}.
\ee
This isomorphism is called blending.

In the fermionic language the Hilbert space $\cH^{(n)}$ lifts to the
Fock space of $n$ fermions $\psi^{(i)}$, $i=1,\ldots,n$, each one with
the expansion (\ref{fermion-NS}) and such that 
\be
\{ \psi^{(i)}_r, \psi^{*\,(j)}_s \} = \delta_{i,j} \delta_{r,-s}.
\ee
Now blending translates to the following redefinitions of these $n$ fermions into a single fermion $\psi$
\be
\psi_{n(r+\rho_i)} = \psi^{(i)}_r,\qquad\qquad \psi^*_{n(r-\rho_i)} = \psi^{*\,(i)}_r,
\ee
where
\be
\rho_i = \frac{2i-n-1}{2n}.  \label{rho-def}
\ee

Blending can also be expressed in terms of two-dimensional partitions introduced above.
Consider $n$ partitions  $R_{(i)}$ of charges $p_i$, with $\sum_i p_i
= p$, corresponding to states in $n$ independent Hilbert spaces of
fermions $\psi^{(i)}$. Associating with each such partition a state of
a chiral fermion $|p_i, R_{(i)} \rangle$, we have a decomposition
\be
|p,{\bf R}\rangle = \bigotimes_{i=1}^n |p_i, R_{(i)}  \rangle,
\ee 
and the blended partition ${\bf R}$ of charge $p$, corresponding to a state in the Hilbert space of the blended fermion $\Psi$, is defined as
\be 
\{ n(p_i+R_{(i),m} - m) +i-1\ |\ m\in\mathbb{N} \} = \{ p+{\bf
  R}_K -K \ | \ K\in\mathbb{N} \}. \label{blend} 
\ee

\section{Some background on $\cD$-modules}\label{sec:dmods}

The theory of $\cD$-modules was introduced and developed, among  others, by
I. Bernstein, M. Kashiwara, T. Kawai and M. Sato, to study linear
partial differential equations from an algebraic perspective \cite{coutinho,bjork,kashiwara,bernstein}.
Currently this is a very active field, with connections and applications to many other branches
of mathematics.

$\cD$-modules are defined as modules for the
algebra of differential operators $\cD$. In general, in a local $\C^n$ patch with complex
coordinates $(z_1,\ldots,z_n)$, the operators $z_i$ and
$\partial_{z_i}$ represent the $n^{th}$ Weyl algebra.
The operators $P\in\cD$ are of the form
\be
P = \sum_{i_1,\ldots,i_n} a_{i_1,\ldots,i_n} \partial_{z_{i_1}}\cdots \partial_{z_{i_n}}.
\ee
With a set of operators $P_1,\ldots,P_m\in\cD$ one can associate a system of differential equations
\be
P_1\Psi = \ldots = P_m\Psi = 0,  \label{PPsi}
\ee 
where $\Psi$ takes values in some function space $\cV$.
An algebraic description of solutions to such a system can be given in terms
of a $\cD$-module $\cM$ determined by an ideal generated by $P_1,\ldots,P_m\in\cD$
\be
\cM = \frac{\cD}{\cD \cdot \langle P_1,\ldots,P_m \rangle}. \label{DDP}
\ee
The advantage of considering such a $\cD$-module is, firstly, that it captures the solutions to the above system
of differential equations independently of the form in which this system is
written. Secondly, it is also independent of the function space $\cV$ -- be it 
the space of square-integrable functions, the space of distributions, the space of holomorphic functions, etc. 

Nonetheless, having chosen a particular space $\cV$ one is interested in, the space of solutions is simply given by the algebra homomorphism
\be
\textrm{Hom}_{\cD} (\cM,\cV).
\ee
E.g. holomorphic solutions to the differential equation $P \Psi(z) =0$
can be captured as a homomorphism of $\cD$-modules
\begin{align}\label{eqn:Dmodule}
\cM =  \frac{\cD}{\cD  \cdot P} ~ \rightarrow~ \cO_{\C},
\end{align}
with $\cO_{\C}$ the algebra of holomorphic functions on the complex plane $\C$. 
Indeed, define a map that sends the element 
\begin{align}
[1] \in \cM ~\mapsto~  \Psi(z) \in
\cO_{\C}. 
\end{align}
This is well-defined because every element $P' \in \cD P$ is mapped to
zero (remember that $\Psi$ fulfills 
$P\Psi=0$), and it is a bijection; conversely, any map $\cM$ to $\cO_{\C}$ is determined by 
a holomorphic solution to the differential equation $P\Psi=0$.

An important notion is a dimension of a $\cD$-module. The so-called Bernstein inequality asserts 
that a non-zero $\cD$-module $\cM$ over the $n^{th}$ Weyl algebra 
has a dimension $2n\geq$ dim$\cM \geq n$. In particular, $\cD$ considered itself as
a $\cD$-module has a dimension $2n$. On the other hand, dim$\,\C[x_1,\ldots,x_n] = n$.
For a non-zero $P\in \cD$, dim$\cD/\cD P = 2n-1$.

A special role in the theory of $\cD$-modules is played by the so-called holonomic $\cD$-modules, which by definition have a minimal dimension $n$. In particular they are cyclic, which means of the form $\{D\Psi: D\in\cD\}$, i.e. they are determined by a single element $\Psi \in\cM$ called a generator.

In the context of the I-brane in $\C^2$ we are just interested in the $1^{st}$ Weyl algebra (\ref{I-Weyl}) of dimension 2.
In this case we immediately conclude that the module $\cD/\cD P$ has a dimension $n=1$ for any non-zero $P$, and is thus holonomic and cyclic. It can be realized as 
\be
\cM = \{D\Psi : D\in\cD\},   \label{DPsi}
\ee
where the generator $\Psi$ is a solution to the differential equation
$P\Psi=0$. 

\subsub{Flat connections}

More generally, $\cD$-modules are defined as differential sheaves on any
variety $X$. The sections of the sheaf $\cD_X$ over an open
neighbourhood $U$ are 
given by linear differential operators on $U$. Therefore, both the
structure sheaf $\cO_X$ (of holomorphic functions) as well as the
tangent sheaf $T_X$ (whose local sections are vector fields) may be
embedded in $\cD_X$
\begin{align}
\cO_X \hookrightarrow \cD_X \hookleftarrow T_X.
\end{align}  
In fact, $\cD_X$ is generated by these inclusions.

A sheaf $\cM$ on $X$ is defined to be a left module for $\cD_X$ when
$v \cdot s \in \cM$, for any  $v \in \cD_X$ and $s\in
\cM$. Furthermore, it has to fulfill  
\begin{align}
v \cdot (fs)& = v(f) s + f(v \cdot s)\\ 
[ v,w ] \cdot s& = v \cdot (w \cdot s) - w \cdot (v \cdot s) \notag
\end{align}
for any $v \in \cD_X$, $f\in \cO_X$ and $s\in \cM$. Suppose that $\cM$
is a left $\cD_X$-module whose sections are the local sections of some
vector bundle $V$ (this encomprises all $\cD_X$-modules that are
finitely generated as $\cO_X$-modules). Then the action of $\cD_X$ defines
a connection on $V$ as
\begin{align}\label{eq:connectioncorrD-mod}
\nabla_v (s) &= v \cdot s, 
\end{align}
whose curvature is zero. So a $\cD$-module structure on the sheaf of
sections of a vector bundle $V$ defines a flat connection on this
vector bundle. And conversely, any module consisting of
sections of a vector bundle $V$ with flat connection $\nabla_A$, has
an interpretation as a $\cD$-module defined through the action of the
flat connection. Therefore, a $\cD$-module is in general just a 
system of linear differential equations, changing from patch to patch on $X$. This is known as a local system. In the main part of this paper $X$ is just $\C$ or $\C^*$.   

\section{Relation to quantum integrable systems}\label{sec:quantuminthier}

In this article we focus on smooth curves that are given by an
equation of the form   
\begin{align}\label{eqn:spectralcurve}
\Sigma: \quad H(z,w) = w^n + u_{n-1}(z) w^{n-1} + \ldots + u_0(z) = 0, 
\end{align}
where $z \in \C$ (or $\C^*$) and $w \in \C$. These play a prominent role
in integrable systems as spectral curves. It is a degree 
$n$ cover over $\C$ (or $\C^*$)
\bea
\Sigma &\subset& T^*\C \notag \\
 \downarrow& \hspace*{-6mm}\pi\\
\C&& \notag
\eea
with possible branch points (from now on we restrict to $z \in \C$ for
simplicity in notation). The spectral curve is embedded in $\C^2$ and  
equipped with the (meromorphic) 1-form 
\bea
\eta = \frac{1}{\la} w dz|_{\Sigma}. 
\eea

Our notion of a quantum
curve agrees with a notion of quantum spectral curves in this
context. Let us say a few words about this.

Fermions on $\Sigma$ transform as holomorphic sections of
a line bundle $L \otimes K^{1/2}_{\Sigma}$, provided by the D6-brane. The pair
$(L, \eta)$ on $\Sigma$ pushes forward to a couple 
\begin{align}\label{eqn:Higgspair}
\pi_*:~ (L, \eta) \mapsto (V = \pi_*L, \phi = \pi_*\eta)
\end{align}
on $\C$ under the projection map $\pi: \Sigma \to \C$. So $V$ is a
rank $n$ vector bundle on $\C$, whereas $\phi$ is a holomorphic 1-form
valued in $gl(n)$.\footnote{In other words, $\phi \in H^0(\C, \mbox{End}V
\otimes K_{\C})$.} Such an object is called a Higgs field. It endows $V$
with the structure of a Higgs bundle. Setting the characteristic polynomial 
\begin{align}
\mbox{det}(\eta - \phi(z))=0
\end{align}
returns the equation for the spectral curve. The push-forward map
$\pi_*$ sets up a bijection between spectral data and 
(stable) Higgs pairs
\begin{align}\label{eqn:Higgsequivalence}
(\Sigma, L) ~\leftrightarrow ~ (V, \phi). 
\end{align}
The moduli space of stable Higgs pairs is an algebraically completely
integrable system, known as the Hitchin integrable system

A $\cD_{\la}$-module (as in \cite{arinkin}) corresponds to a
$\la$-connection~$\nabla_{\la}$  
\begin{align}
\nabla_{\la} = \la \partial_z - A(z),
\end{align}
which is defined through the Leibnitz rule $\nabla_{\la} (fs) = f
\nabla_{\la}(s) + \la s \otimes df$ for any function $f$ and section $s$.

Semi-classically, such a $\la$-connection $\nabla_{\la}$ reduces to a
1-form $\nabla_0(z)$ with values in $gl(n)$  
\begin{align}
\nabla_{\la} \mapsto \nabla_0, \quad (\la \to 0).
\end{align}
We just encountered this object as a Higgs field $\phi$. Moreover, we
explained with (\ref{eqn:Higgspair}) that a Higgs $(V,\phi)$ and
spectral data $(\Sigma,L)$ 
provide equivalent information. In particular, the spectral curve can
be recovered by the determinant of the Higgs field. This implies that
$\la$-connections quantize spectral data.\footnote{These
  $\la$-connections are also known as $\la$-opers, and play an important
  role in the quantum integrable system of Beilinson and Drinfeld
  \cite{beilinsondrinfeld,talkarinkin}. }

It tells us exactly which requirements a $\cD$-module quantizing the
I-brane configuration has to satisfy. Fermions on 
a degree $n$ spectral curve have to transform under a rank $n$
$\la$-connection $\nabla_{\la}$ on $\C$, whose semi-classical $\la \to 0$
limit is given by the Higgs field 
\begin{align}
\nabla_0 = \pi_* (\eta).
\end{align}
A simple
example of a $\la$-connection is given by  
\begin{align}
\nabla = \la \partial_z - A(z),
\end{align}
with $A(z) = \pi_* (\eta)$. Its determinant is a degree $n$
differential equation that canonically quantizes the defining equation
for $\Sigma$.


{\small

\end{document}